\documentclass[aps,pra,twocolumn,floatfix,superscriptaddress,
showkeys,amsmath,amssymb,longbibliography]{revtex4-2}		
\usepackage{braket}
\usepackage{graphicx}
\usepackage{dcolumn}
\usepackage{bm}
\usepackage[colorlinks,urlcolor=blue,citecolor=blue,linkcolor=blue]{hyperref}
\usepackage{subfig}
\usepackage[font=small,labelfont=bf,format=plain,justification=centerlast,labelsep=period]{caption}
\usepackage{tikz}
\usetikzlibrary{arrows,shapes,positioning,shadows,svg.path,backgrounds,fit}

\usepackage{verbatim}
\usepackage[normalem]{ulem}
\usepackage{orcidlink}

\newcommand{\mi}{\mathrm{i}}

\begin{document}


\title{Spectral response of a nonlinear Jaynes-Cummings model}

\author{L. Medina-Dozal\,\orcidlink{0000-0002-4695-5190}}
\author{A.R. Urz\'ua\,\orcidlink{0000-0002-6255-5453}}
\author{D. Aranda-Lozano}
\author{C. A. Gonz\'alez-Guti\'errez\,\orcidlink{0000-0002-1734-1405}}
\author{J. R\'ecamier\,\orcidlink{0000-0002-5995-0380}}
\affiliation{
Instituto de Ciencias F\'isicas, Universidad Nacional Autónoma de M\'exico. Avenida Universidad s/n, Col. Chamilpa, Cuernavaca, Morelos, 62210 Mexico
}
\author{R. Rom\'an-Ancheyta\,\orcidlink{0000-0001-6718-8587}}
\email{ancheyta@fata.unam.mx}
\affiliation{
Centro de F\'isica Aplicada y Tecnolog\'ia Avanzada, Universidad Nacional Aut\'onoma de M\'exico, Boulevard Juriquilla 3001, Querétaro 76230, Mexico
\\}

\date{\today}

\begin{abstract}
The Jaynes-Cummings quantum optics model allows us to understand the dialogue between light and matter at its most fundamental level\textcolor{black}{, which is crucial for advancements in quantum science and technology}. Several generalizations of the model have long been proposed, emphasizing their dynamic behavior but paying less attention to their spectroscopy. Here, we obtain analytical expressions of the time-dependent spectral response of a nonlinear \textcolor{black}{Jaynes-Cummings} model based on deformed field operators. We show that the long-time response of the resulting nonlinear cavity field resembles the one experimentally obtained in the strong-dispersive regime of circuit quantum electrodynamics. The spectrum is intrinsically asymmetric with the nonlinear coupling, a signature of the impossibility of getting resonant conditions for finite field excitations.
\end{abstract}

\maketitle

\section{Introduction}

One of the most valuable approaches to understanding quantum systems' physical behavior is through their spectral response~\cite{Ficek_2017}.
This has been the case in atomic, molecular, and optical physics, superconducting quantum circuits, quantum materials, and any other field that uses electromagnetic radiation to measure matter properties. For instance, in the polaritons panorama~\cite{Asenjo_2021}, the display of the spectral Rabi splitting and its asymmetries is evidence, respectively,
for the hybridization of light-matter quantum states~\cite{Kasprzak_Nat_Matt_2010} and as a confirmation of new avenues of light-matter interaction, the so-called ultra-strong~\cite{Nori_2019} and deep-strong~\cite{Reich_2020}  coupling regimes 
 
The time-dependent physical spectrum is a powerful tool for obtaining the spectral response of quantum systems. It was proposed by Eberly and W\'odkiewicz (EW) in~\cite{Eberly1977} and is defined as follows,
\begin{equation}\label{EWspec}
\begin{aligned}
	{S}(\omega,\!\Gamma,\!t)\!=\! 
	2\Gamma e^{-2\Gamma t}\!\! \int_{0}^{t}\! dt_{1}^{}\!\!\int_{0}^{t}\!\!dt_{2}^{} e^{(\Gamma - \mathrm{i}\omega)t_{1}} e^{(\Gamma + \mathrm{i}\omega)t_{2}}G(t_{1},\!t_{2}),
\end{aligned}
\end{equation}
where $G(t_{1}, t_{2})$ is the two-time autocorrelation function computed in respect to some given initial state. In contrast to the stationary Wiener-Khintchine power spectrum, Eq.~(\ref{EWspec}) can track the spectral evolution of non-stationary systems (such as those considered in this work), and it is always positive since it is based on a fully quantum description of the photodetection process. Due to the nonzero bandwidth filter $\Gamma$, the time-energy uncertainty is appropriately handled. Recently, the EW spectrum has proven to be useful in diverse scenarios, such as the quantum thermometry of non-thermal baths~\cite{Ancheyta_2020}, intermittent resonance fluorescence~\cite{Ancheyta_2018}, ultrastrong coupling~\cite{Ancheyta_2021}, and even the strain-spectroscopy of superconducting qubits~\cite{Ancheyta_2022}.

In this work, we obtain the Eberly-W\'odkiewicz spectrum of a nonlinear Jaynes-Cummings (JC) model, the deformed JC model~\cite{Santos2012, Cordero}, where the corresponding Hamiltonian is based on the f-oscillators introduced in~\cite{manko1} generalizing the q-oscillators. \textcolor{black}{These are nonlinear oscillators, i.e., their equations of motion are nonlinear, their frequency depends on the energy~\cite{Manko93, Bhattacharjee_07}, and they can even be used to harvest work in exotic quantum thermal machines~\cite{PRE_Ozgur_23}.} In~\cite{manko1}, the usual annihilation ($\hat{a}$) and creation ($\hat{a}^\dagger$) boson field operators transform to $\hat{A}$ and $\hat{A}^{\dagger}$ in a noncanonical way, where $\hat{A} = \hat{a}f(\hat{n})$, $\hat{A}^{\dagger} = f(\hat{n})\hat{a}^{\dagger}$, being $f(\hat{n})$ the so-called deformation function, a function of the number operator $\hat{n}=\hat{a}^\dagger\hat a$. This function introduces \textcolor{black}{nonlinearities} in bare Hamiltonians and intensity-dependent couplings in interacting systems that can naturally emerge, for instance, in the physics of nonlinear potentials~\cite{Ancheyta_2011}, trapped ions~\cite{Vogel,Moya_Reports12,PRA_Solano_18}, excitons in \textcolor{black}{semiconductor} quantum dots~\cite{Liu_PRA_2001, Harouni_JPB_2009, Laussy_PRB_2006}, or photonic lattices~\cite{Blas_OpExp_2013}.

Here, we show in detail $i)$ how the spectral response of an f-oscillator strongly resembles the one experimentally obtained in the strong-dispersive regime of circuit quantum electrodynamics (QED)~\cite{Schuster2007,PRA_Palmer15}
 and $ii)$  that the response of the deformed JC model is intrinsically asymmetric due to the impossibility of getting resonant conditions for arbitrary field excitations, a direct consequence of the nonlinear field and the nonlinear interaction. Unless $f(\hat n)\!=\!1$, the commutator $[\hat{A},\hat{A}^\dagger]\neq 1$, therefore, our spectroscopic results contribute to the recent experimental~\cite{Alderete_2021, Alderete_2022} and theoretical~\cite{Alderete_PRA_2017, Alderete_2018} efforts for the quantum simulation of para-particles oscillators, i.e., particles with statistics that deviate from the Bose-Einstein and Fermi-Dirac distributions.  

Since we want to make our article as self-contained as possible, we organize it as follows. In Sec.~\ref{Section_2}, we briefly review the solution of the standard Jaynes-Cummings model and use it to obtain the corresponding autocorrelation function to compute the EW spectrum. We describe the abovementioned approach in some detail, as it provides the basis for most of the following. Section~\ref{Section_3} is devoted to the spectral analysis of the deformed JC model, specifically, Sec.~\ref{Section_Nonlinear_Oscillator} gives the EW spectrum of the corresponding nonlinear field in the bare Hamiltonian, and Sec.~\ref{Section_Nonlinear_Coupling} treats the entire system, including the nonlinear coupling. We give our conclusions in Sec.~\ref{Section_4}.
\textcolor{black}{Appendices~\ref{app:corrfunc} and \ref{app:parity} detail the essential steps to obtain the autocorrelation function, the EW spectrum, and the parity preservation of the JC model.}

\section{Standard Jaynes-Cummings model}\label{Section_2}

In the usual treatment of a two-level atom coupled to a quantized electromagnetic field, under the rotating-wave approximation (RWA), the Hamiltonian can be expressed as~\cite{Jaynes_1963,Haroche2006,Liberato2024}
$\hat{H}_{\texttt{JC}}= \hbar\omega_{c}(\hat a^\dagger \hat a+\hat a\hat a^\dagger)/2 + \hbar{\omega_{a}}\hat{\sigma}_{z}/2 - \mathrm{i}\hbar{\Omega_{0}}\left(\hat{a}\hat{\sigma}_{+} - \hat{a}^{\dagger}\hat{\sigma}_{-}\right)/2$, where $\omega_{c}$ and $\omega_{a}$ are the frequencies of the cavity field and the two-level system respectively, $\Omega_{0}$ is the vacuum Rabi frequency which measures the coupling strength  between the field and the atomic system. Here, $\hat a$ and $\hat a^{\dagger}$ satisfy $[\hat a,\hat a^{\dagger}] =1$ and  $\hat\sigma_+$, $\hat \sigma_{-}$ are the atomic ladder operators with commutator $[\hat \sigma_{+},\hat \sigma_{-}]=\hat \sigma_z$. The eigenstates for the uncoupled ($\Omega_0=0$) system are the tensor product of the states $|e,n\rangle\equiv|e\rangle\otimes|n\rangle$ and $|g,n+1\rangle\equiv|g\rangle\otimes|n+1\rangle$ of atomic and cavity energy eigenstates. 
Since the interaction only generates transitions when the field gains an excitation, and the atom goes from the excited to the ground state and vice versa, $\ket{e, n} \rightarrow \ket{g, n + 1}$  and $\ket{g, n + 1} \rightarrow \ket{e, n}$, the total number of excitations is a conserved quantity, and one can interpret the Hamiltonian as that of $n$ independent two-level systems. Only the ground state $\ket{g, 0}$ is an unpaired state and does not couple with any \textcolor{black}{other} state.

Then, the solution can be obtained by the diagonalization of the $n-$th doublet which contains $n + 1$ excitations, either $n$ from the field and one from the atom, or  $n + 1$ from the field and no excitation from the atom. In this basis, the matrix representation of the Hamiltonian (without the zero-point energy $\hbar\omega_c/2$) is
\begin{equation}\label{hjc_diag}
    {H}_n = \hbar\omega_{c}\big(n+{1}/{2}\big){\mathbb{I}}_{2\times 2}^{}+V_{n},\,\,V_{n} = \frac{\hbar}{2}\begin{pmatrix}
		\Delta & -\mathrm{i}\Omega_{n}\\
		\mathrm{i}\Omega_{n} & -\Delta
	\end{pmatrix}\!,
\end{equation}
where ${\mathbb{I}}$ is the identity matrix, $\Delta = \omega_{a} - \omega_{c}$ is the atom-field detuning and $\Omega_n=\Omega_0\sqrt{n+1}$ is the $n$-photon Rabi frequency.  
Diagonalization of \eqref{hjc_diag} yields the eigenvalues ${E_{n}^{\left(\pm\right)}} = {\hbar}\omega_{c}(n + {1}/{2}) \pm {\hbar}\sqrt{\Delta^{2} + \Omega_{n}^{2}}/2$~\cite{Haroche2006}, whose eigenvectors are
$|{+, n}\rangle= \cos(\theta_{n}/2)\ket{e, n} + \mathrm{i}\sin(\theta_{n}/2)\ket{g, n + 1}$,
and
$|{-, n}\rangle= \sin(\theta_{n}/2)\ket{e, n} - \mathrm{i}\cos(\theta_{n}/2)\ket{g, n + 1}$, called dressed states,
with the mixing angle $\tan \theta_n =\Omega_n/\Delta.$ \ 
At resonance ($\Delta = 0$), the mixing angle is $\theta_{n} = \pi/2$ 
so the dressed states become
$\ket{\pm, n} = \left(\ket{e, n} \pm \mathrm{i}\ket{g, n + 1}\right)/\sqrt{2}$.
Given an initial state written in terms of the uncoupled basis, one can go to the dressed states and evolve the system to obtain the state at later times. 

As an illustrative example, consider the scenario where the initial state is $\ket{\Psi(0)} = \ket{e, n}$ and $\Delta=0$, written in terms of the dressed states we have $\ket{e, n} = \left(\ket{+, n} + \ket{-, n}\right)/\sqrt{2}$.
The evolution of $|\Psi(0)\rangle$ \textcolor{black}{using the full Hamiltonian yields
\begin{equation}
\begin{aligned}
&|\Psi(t)\rangle\!=\!e^{-i\omega_a(\!n+1/2\!)t}\big(e^{-\mathrm{i}\Omega_{n}t/2}\!\ket{+, n}\!+\!e^{\mathrm{i}\Omega_{n}t/2}\!\ket{-, n}\big)\!/\!\sqrt{2},\\
    &=e^{-i\omega_a(\!n+1/2\!)t}\big[\!\cos\left(\Omega_n t/2\right)|e, n\rangle\!+\!\sin\left(\Omega_n t/2\right)|g, n\!+\!1\rangle\big]. 
\end{aligned}
\end{equation}
}
To obtain the EW spectrum, it is necessary to compute the two-time autocorrelation function $G(t_{1}, t_{2}) = \braket{\hat{O}^{\dagger}(t_{1})\hat{O}(t_{2})}$ regarding some initial state. $\hat{O}$ and $\hat{O}^\dagger$ are the ladder operators of the system of interest in the Heisenberg representation~\cite{Ancheyta_2011}. For the two-level atom, the correlation is given by
\begin{equation}\label{func:acorr}
     G(t_{1},\!t_{2})_{\rm atom}^{}\!=\! 
    \braket{\Psi(0)\vert\hat{U}^{\dagger}\!(t_{1}\!)\hat{\sigma}_{+}\hat{U}(t_{1}\!) \hat{U}^{\dagger}(t_{2}\!)\hat{\sigma}_{-}\hat{U}(t_{2}\!)\vert\Psi(0)}\!, 
\end{equation}
\textcolor{black}{where $\hat U(t_j)=\exp(-{\rm i} \hat H_{\texttt{JC}}t_j/\hbar)$ is the time-evolution operator.} 
Since $|\Psi(0)\rangle = |e,n\rangle$ is our initial state, the correlation reduces to  \textcolor{black}{(see Appendix \ref{app:corrfunc})}:
\begin{equation}\label{func:atom_corr}
\begin{aligned}
    G(t_{1}, t_{2})_{\mathrm{atom}}^{} & = e^{\mathrm{i}\omega_{a}(t_{1} - t_{2})}\cos({\Omega_{n}}t_{1}/2)\\ 
    &\times\cos[{\Omega_{n - 1}}(t_{1}-t_{2})/2]\cos({\Omega_{n}}t_{2}/2), 
\end{aligned}
\end{equation}
from where we can obtain the time-dependent spectrum as defined in~(\ref{EWspec}).
When $n=0$, it is straightforward to show that the EW physical spectrum in the long-time limit ($\Gamma t\gg 1$) is~\cite{Sanchez_PRL_1983, H_Paul_1986}
\begin{equation}\label{Slimit_nondef}
\small
S(\omega,\Gamma)_{\texttt{VRS}}^{}=\frac{{\Gamma}/{2}}{\Gamma^2+\big(\omega-\omega_a+\frac{\Omega_0}{2}\big)^2}+\frac{\Gamma/2}{\Gamma^2+\big(\omega-\omega_a-\frac{\Omega_0}{2}\big)^2},
\end{equation}
where VRS means vacuum Rabi splitting. The distance between the two Lorentzian is $\Omega_0$. As expected, when $\Omega_0\rightarrow 0$, the atomic spectral response is a single Lorentzian peak centered at the atomic transition frequency $\omega_a$. It is worth noting that since the JC Hamiltonian reproduces the lowest energy sector of the quantum Rabi one, see Ref.~\cite{Liberato2024} and also Fig.~\ref{fig:eigenvals}~$a)$, we may consider Eq.~(\ref{Slimit_nondef}) as a good approximation of the VRS, but in the ultrastrong coupling regime with large values of the normalized coupling constant up to $\Omega_0/2\omega_a\sim 0.3$. The Hamiltonian of the quantum Rabi model is $\hat{H}_{\texttt{Rabi}}=\hbar\omega_c\hat a^\dagger\hat a+\hbar\omega_a\hat \sigma_z/2-\mathrm{i}\hbar\Omega_{0}(\hat{a}-\hat{a}^{\dagger})\hat\sigma_x/2$.

Now, we turn our attention to the counterpart when we see the field correlation function. This aims to observe the long-time behavior. The two-time correlation function corresponding to the field is $G(t_{1}, t_{2})_{\mathrm{field}}^{}=\braket{\Psi(0)\vert\hat{a}^{\dagger}(t_{1})\hat{a}(t_{2})\vert \Psi(0)}$.

Considering the same initial state as before, $\ket{\Psi(0)}=|e,n\rangle$, we obtain
\begin{equation}
\begin{aligned}
    &G(t_{1}, t_{2})_{\mathrm{field}} = \frac{e^{\mathrm{i}\omega_{c}(t_{1} - t_{2})}}{4}\left[\phantom{\cos\big(\frac{\Omega_{-}}{2}(t_{1} - t_{2})\big)}\right.\\
    &\left.\Big(1 + 2n + 2\sqrt{n(n + 1)}\Big)\cos\left(\frac{\Omega_{-}}{2}(t_{1} - t_{2})\right)\right.\\
    &\left.-\cos\left(\frac{1}{2}(\Omega_{-}t_{2} + \Omega_{+}t_{1})\right) - \cos\left(\frac{1}{2}(\Omega_{-}t_{1} + \Omega_{+}t_{2})\right)\right.\\
    &\left.+\left(1 + 2n - 2\sqrt{n(n+1)}\right)\cos\left(\frac{\Omega_{+}}{2}(t_{1} - t_{2})\right)\right],
\end{aligned}
\end{equation}
where $\Omega_{+} = \Omega_{n} + \Omega_{n - 1}$ and $\Omega_{-} = \Omega_{n} -\Omega_{n - 1}$. The field correlation function appears to be more involved than the atomic one. However, a similar expression to Eq. \eqref{Slimit_nondef} can be obtained for the physical spectrum in the case of $n=0$, i.e., the vacuum Rabi splitting is insensitive to the quantum statistics~\cite{H_Paul_1986, Laussy_PRB_2006}. When $\Omega_0=0$, it is easy to show that the field's spectral response is also a single Lorentzian centered at the field's frequency, $\omega_c$, whose location is independent of its initial state. This means we recover the usual spectral response of a single quantum harmonic (linear) oscillator~\cite{Ancheyta_2020}. Such a picture is drastically affected, as shown in the next section when we consider the \textcolor{black}{nonlinearities} of $f(\hat n)$.

\section{Deformed Jaynes-Cummings model}\label{Section_3}

The standard Jaynes-Cummings model has been the subject of important generalizations, for example, incorporating a nonlinear coupling between the atom and the field \cite{Villamizar_2021,Buck,Buzek,Cordero,Yang}, incorporating a group of two-level atoms in the cavity~\cite{Tavis}, and also incorporating nonlinear terms in the photon number operator to explore the time evolution of the system in a Kerr-like medium \cite{Agarwal,Gora,Xie}. However, \textcolor{black}{with some relevant exceptions~\cite{Buzek1990,Feng2000},} little has been said about its spectral response.

Following the ideas introduced in~\cite{manko1} in this section we consider the deformed operators $\hat{A}$, $\hat{A}^{\dagger}$ and the function $f(\hat{n})$. As we mentioned in the introduction, $f(\hat{n})$ is responsible for the appearance of \textcolor{black}{nonlinearities} in the system Hamiltonian. For instance, if $f^2(\hat{n}) =(\hbar a^2/2\mu\Omega)(2s + 1 - \hat{n})$ the deformed Hamiltonian $\hat{H}_{f} = {\hbar\Omega}(\hat{A}^{\dagger}\hat{A} + \hat{A}\hat{A}^{\dagger})/2$ becomes $\hat{H}_{f} = ({\hbar^{2}a^{2}}/{2\mu})(s+2s\hat{n}-\hat{n}^{2})$, whose energy spectrum is identical to that of the modified P\"oschl-Teller potential $V(x)=U_0\tanh^2(ax)$~\cite{santos2011}, where $a$ represents the range of the potential, $\mu$ is the reduced mass, and $s$ is related to the well's depth by $s(s+1)=2\mu U_0/(\hbar^2a^2)$. This anharmonic potential successfully describes molecular vibrations~\cite{LEMUS2002401} and, recently, the $\rm{CO_2}$ experimental Raman spectra~\cite{Lemus_2020}. 

For an arbitrary function, $f(\hat n)$ the commutator between $\hat A$ and $\hat A^{\dagger}$ is
$[\hat{A}, \hat{A}^{\dagger}] = (\hat{n} + 1)f^2(\hat{n} + 1) - \hat{n}f^2(\hat{n})$,
a function of the number operator, and it can be seen that it reduces to one when $f(\hat n)$ goes to one. The commutator of the deformed operators with the number operator $\hat n$ is $[\hat A,\hat n] = \hat A, \ \ [\hat A^{\dagger},\hat n]= -\hat A^{\dagger}$, which is the same as that of the usual creation-annihilation operators with the number operator.

The deformed \textcolor{black}{(nonlinear)} Jaynes-Cummings Hamiltonian in the rotating-wave approximation is~\cite{Santos2012,Blas_OpExp_2013}
\begin{equation}\label{eq:hdef}
    \hat H^{D}_{\texttt{JC}}\!=\!\frac{\hbar\omega_{c}}{2}(\hat{A}^{\dagger}\hat{A}+\hat{A} \hat{A}^{\dagger}) + \frac{\hbar\omega_a}{2}\hat{\sigma}_{z}\!-\mathrm{i}\frac{\hbar\Omega_{0}}{2}(\hat{A}\hat{\sigma}_{+}\!-\!\hat{A}^{\dagger} \hat{\sigma}_{-}),
\end{equation}
which written in terms of $\hat a$, $\hat a^\dagger$ and $\hat n$, yields
\begin{equation}\label{deform_JCH}
\begin{aligned} 
    \hat H^{D}_{\texttt{JC}} &= \frac{\hbar\omega_{c}}{2}\big[\hat{n} f^{2}(\hat{n})+(\hat{n} + 1)f^{2}(\hat{n} + 1)\big] + \frac{\hbar\omega_a}{2}\hat{\sigma}_{z}\\
    &\qquad\quad -{\rm i}\frac{\hbar}{2}\hat{a}\hat{\sigma}_{+}^{}\big[{\Omega_0f(\hat{n})}\big] +{\rm i}\frac{\hbar}{2}\big[{\Omega_0f(\hat{n})}\big]\hat{a}^{\dagger}\hat{\sigma}_{-}^{}.
\end{aligned}
\end{equation}
\textcolor{black}{We refer to the deformed JC model as a nonlinear JC model because, in addition to involving deformed operators associated with a nonlinear oscillator with an intensity-dependent frequency (see Sec.~\ref{Section_Nonlinear_Oscillator}), Hamiltonian~(\ref{deform_JCH}) shows the intensity-dependent coupling $\Omega_0f(\hat n)$.}
Before entering into the complete study of the total deformed JC model, we first examine the spectral response given by the bare Hamiltonian of Eq.~(\ref{eq:hdef}). In particular, we analyze the field contribution since it is the part that depends on the deformation function; see Eq.~(\ref{deform_JCH}).

\subsection{Spectral response of a nonlinear field}\label{Section_Nonlinear_Oscillator}

\begin{figure}[t]
     \centering
     \includegraphics[width= \linewidth, keepaspectratio]{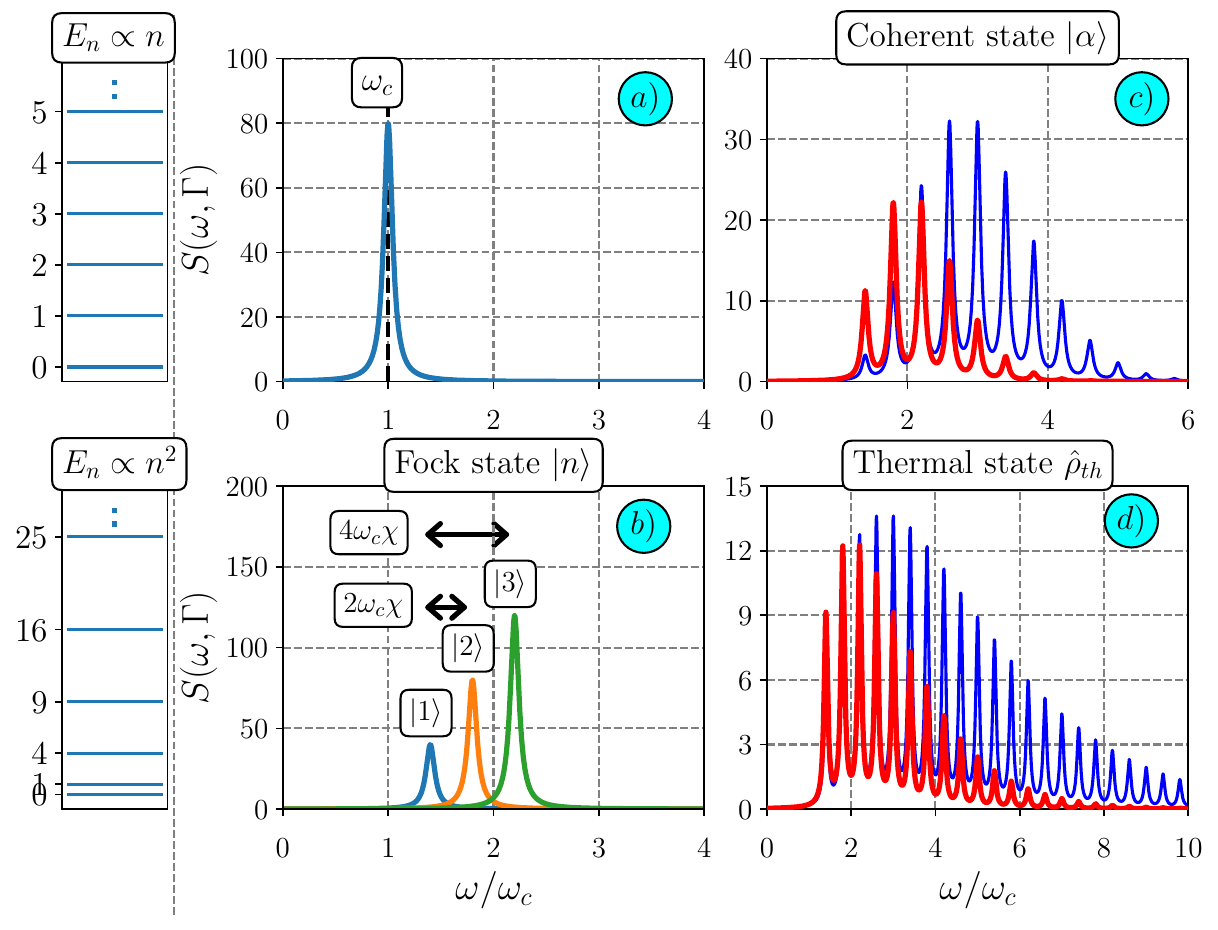}     
     \caption{Top and bottom left: Energy level structure of a linear (nonlinear) cavity field. $a)$ Spectral response of a linear field. The Lorentzian is centered at frequency $\omega_c$, independent of the initial state. $b)$ Spectral responses of the nonlinear field. The frequency and amplitude of each Lorentzian depends on the initial Fock state of the field; see Eq.~(\ref{EW_nonlinear_oscillator}). $c)$ and $d)$ Same as $b)$ but with coherent and thermal initial states, respectively; see Eq.~(\ref{SW_nonlinear_coherent}). The solid red (blue) line is for an average photon number $\bar{n}=2$ $(4)$. The rest of the parameters are $\chi = 0.2$, and $\Gamma = 0.05\omega_c$}
\label{EW_Nonlinear_Field_2}
 \end{figure}

\textcolor{black}{In this article, we decided to work with the simple} deformation function $f^2(\hat{n})=1+\chi\hat{n}$~\cite{Cordero, Ancheyta_2014}, \textcolor{black}{which allows us to introduce} a nonlinear coupling between the two-state atom and the field and a Kerr-\textcolor{black}{type nonlinearity}, where $\chi$ is a dimensionless small parameter. The field Hamiltonian $\hat H_{f}\!=\!{\hbar\omega_{c}}(\hat{A}^{\dagger}\hat{A}\!+\!\hat{A} \hat{A}^{\dagger})/2$ is
\begin{equation}\label{Nonlinear_Field_fn}
\hat H_{f} =\hbar\omega_c(\hat n+1/2)+\hbar\omega_c\chi(\hat n^2+\hat n+1/2).
\end{equation} 
It has a clear nonlinear energy spectrum proportional to $n^2$, which can be easily emulated in photonic lattices with quadratic refractive index profile~\cite{Blas_OpExp_2013, Ancheyta_PRA_2017}, as well as in superconducting circuits using a transmon qubit (see Sec.~\ref{circuit_implementation}). 
\textcolor{black}{We note that more complicated but genuinely physical deformation functions exist in the literature. For instance, for the {\it q}-oscillators we have~\cite{manko1, Harouni_JPB_2009} 
\begin{equation}\label{f_qoscillators}
      f(\hat n)=\sqrt{\frac{1}{\hat n}\frac{q^{\hat n}-q^{-\hat n }}{q-q^{-1}}}=\sqrt{\frac{\sinh \lambda\hat n}{\hat n \sinh \lambda}},
\end{equation}
where $\lambda=\log q$ and $q$ is a real parameter in the ``q-commutator'' $\hat a_q\hat a_q^\dagger-q\, \hat a_q^\dagger \hat a_q=1$. For high-density Frenkel excitons (electron-hole pairs), the {\it q}-parameter is not phenomenological and depends on the total molecule number ($N$) as $q=1-2/N$~\cite{Liu_PRA_2001}. Therefore, when $N\gg 1$, $q\sim 1$ and $\lambda$ is close to zero. In this situation, we can make a power series expansion of Eq.~(\ref{f_qoscillators}) about the point $\lambda=0$ to second order and approximate the square of the deformation function as $f^2(\hat n)\approx 1+\frac{2}{3N^2}(\hat{n}^2-1)$.
In ion-traps, the deformation function is~\cite{Vogel,Moya_Reports12,PRA_Solano_18}
\begin{equation}\label{f_ions}
f(\hat{n})=e^{-\eta^2/2}\sum_{l=0}^\infty\frac{(-\eta^2)^l}{l!(l+1)!}\frac{\hat n!}{(\hat n-l)!},
\end{equation}
where $\eta$ is the Lamb-Dicke parameter that measures the vibrational amplitude of the ion. Near the Lamb-Dicke regime $\eta$ is a small number, and the vibrational amplitude of the ion is much smaller than the wavelength of the laser field. In this situation, a power expansion of (\ref{f_ions}) yields $    f^2(\hat n)\approx 1-\eta^2(\hat n+1)$, which is similar to the linear dependency of the number operator we are using.
}
\textcolor{black}{Hamiltonian (\ref{Nonlinear_Field_fn})} generates the time-evolution $\hat a(t)=\hat a(0)\exp[-\mathrm{i}(\omega_c+2\omega_c\chi \hat n)t]$. In obtaining $\hat a(t)$, we used the identity $\hat a f(\hat n)=f(\hat n+1)\hat a$ and its adjoint. Notice how, due to the Kerr nonlinearity, the phase now depends explicitly on the number operator~\cite{Ancheyta_2017}; if $\chi\rightarrow 0$, we recover the harmonic limit (Heisenberg-Weyl symmetry). For an initial Fock state $|n\rangle$ the two-time autocorrelation function of the nonlinear cavity field takes the form 
\begin{equation}
     G_n(t_1,t_2) = n\exp[\mathrm{i}(\omega_c+2\omega_c\chi n)(t_1-t_2)],
\end{equation}
from which the spectrum at any time can be obtained, while in the long-time limit ($\Gamma t\gg 1$) the field's physical spectrum (\ref{EWspec}) is
\begin{equation}\label{EW_nonlinear_oscillator}
    S_n(\omega,\Gamma)^{}=\frac{2 n \Gamma}{\Gamma^2+(\omega-\omega_c-2\omega_c\chi n)_{}^2}.
\end{equation}
In contrast to the quantum harmonic oscillator, Eq.~(\ref{EW_nonlinear_oscillator}) is a Lorentzian whose frequency center depends on the field excitation; it is \textcolor{black}{blue-}shifted by $2\omega_c\chi n$; see Fig.~\ref{EW_Nonlinear_Field_2}~$b)$. This is an interesting, unreported, and by no means a trivial result; it represents the simplest spectroscopic response of a nonlinear quantum field in the most quantum state, the Fock state. Its classical analog is straightforward; it corresponds to an anharmonic oscillator whose frequency and amplitude are functions of the energy~\cite{Bhattacharjee_07, Sivakumar_2000}. Our latter observation is crucial to understand the minimum uncertainty coherent-states formalism for general potentials using noncanonical transformations~\cite{Nieto_PRL_1978, Nieto_PRD_I, Nieto_PRD_II}. 

On the other hand, if the field described by $\hat H_f$ starts in a coherent state $|\alpha\rangle$,  with $|\alpha|^2=\bar{n}$ the average photon number, the two time autocorrelation function is $G_{\bar n}^{}(t_1,t_2)=\sum_n^{} P_n^{}(\bar{n}) G_n^{}(t_1,t_2)$ where $P_n(\bar{n})=e^{-\bar{n}}\bar{n}^n/n!$ is the Poisson probability distribution. In the long-time limit, the corresponding physical spectrum is 
\begin{equation}\label{SW_nonlinear_coherent}
    S_{\bar n}^{}(\omega,\Gamma)={\sum}_n^{} P_n^{}(\bar{n})S_n^{}(\omega,\Gamma).
\end{equation} 
A classical intuition might tell us that when the system is in a coherent state, which is the most classical state, we should recover a classical behavior (a single spectral peak). 
Intuition, in this case, fails, $S_{\bar n}^{}(\omega,\Gamma)$ corresponds to a sum of Lorentzians (multi-peak structure) weighted by the photon number distribution $P_n(\bar{n})$; see Fig.~\ref{EW_Nonlinear_Field_2}~$c)$. When $\chi\rightarrow 0$, $S_{\bar n}^{}(\omega,\Gamma)$ reduces to the well-known spectral response of the harmonic oscillator with a single central frequency and amplitude $\propto\bar{n}$; see Fig.~\ref{EW_Nonlinear_Field_2}~$a)$.

When the initial state of the field is the thermal state $\hat{\rho}_{\rm th}=\sum_n P_n^{\rm th}(\bar{n})|n\rangle\langle n|$, the corresponding spectral response is $S_{\rm th}^{}(\omega,\Gamma)=\sum_n^{} P_n^{\rm th}(\bar{n})S_n^{}(\omega,\Gamma)$, where $P_n^{\rm th}(\bar{n})=\bar{n}^n/(1+\bar{n})^{n+1}$ and $\bar{n}=[\exp(\hbar\omega_c/k_BT)-1]^{-1}$ is the Bose-Einstein distribution.  

Figure~\ref{EW_Nonlinear_Field_2} shows (left panel) a schematic representation of the equally spaced eigenvalues corresponding to a linear system and the unequally spaced energy levels corresponding to the nonlinear energy spectrum. In Fig.~\ref{EW_Nonlinear_Field_2}~$a)$, we show the EW spectrum for a number state and a linear field; there is one peak centered at the frequency $\omega_c$, which is independent of the initial state~\cite{Ancheyta_2020}. In Fig.~\ref{EW_Nonlinear_Field_2}~$b)$, we notice the effects due to the nonlinearities in the field, displacing the peaks towards larger frequencies corresponding to different number states. The position and amplitude of the peaks become a function of the number state. In Fig.~\ref{EW_Nonlinear_Field_2}~$c)$, the EW spectrum for a coherent state with an average number of photons $|\alpha|^2=2$ (red) and $|\alpha|^2=4$ (blue), and we observe a displacement of the peaks to larger frequencies in accord with $b)$. In $d)$, we show the EW spectrum for a thermal state with an average number of photons $\bar n=2$ (red) and $\bar n=4$ (blue)
 
Interestingly, the three different spectral responses of the nonlinear field shown in Fig.~\ref{EW_Nonlinear_Field_2} strongly resemble the experimental ones of an artificial two-level atom in the strong-dispersive regime of a superconducting circuit~\cite{Schuster2007}\textcolor{black}{; see also~\cite{PRA_Palmer15}}. In such a regime ($\Delta\gg \Omega_0$) the standard JC interaction reduces to the dispersive interaction $\hat H_{\rm disp}=\hbar(\Omega_0^2/4\Delta)\hat n\hat\sigma_z$~\cite{gerry_knight_2004}. $\hat H_{\rm disp}$ commutes with the atomic and field bare Hamiltonians of the JC model. Although the qubit cannot absorb photons from the resonator (field), Ref.~\cite{Schuster2007} demonstrated, how the qubit transition energy can be spectrally resolved into different spectral lines similar to the ones presented in Fig.~\ref{EW_Nonlinear_Field_2}. In addition, our analytical expressions allow us to investigate the system's spectral response for arbitrarily high photon numbers. This contrasts with the limited numerical simulation in~\cite{Schuster2007}, where only average photon numbers $\bar{n}<4$ were used. Our results confirm from a spectroscopic standpoint that the field Hamiltonian (\ref{Nonlinear_Field_fn}) behaves, indeed, as a Kerr-like medium with the ``self-interaction'' nonlinear term $\hbar\omega_c\hat \chi \hat n\hat n$, which commutes with the linear one $\hbar\omega_c\hat n$. 

\textcolor{black}{
We want to stress the role of the filter bandwidth ($\Gamma$) in the spectral responses described above. Since the EW spectrum is rooted in experimental conditions of photodetection, any spectrally resolved detected light field is inevitably influenced by a nonzero $\Gamma$. For a Fabry-Perot interferometer acting as the filter placed before a broadband detector, $\Gamma^{-1}$ is its filling time and depends on the distance between the plates and the reflection coefficient~\cite{Eberly1977}. In Fig.~\ref{EW_Nonlinear_Field_2} and throughout the article, we use feasible values to remain within the narrow-band detection limit, i.e., $\Gamma/\omega_c\ll \chi$ and  $\Gamma/\Omega_0\ll 1$~\cite{PRL_Finley_09, Nazir_NatP_2017, Tilmann23}. Larger values of $\Gamma$ reduce spectral resolution, causing a departure from the strong-dispersive and the strong-coupling regimes. Unlike the unrealistic infinite spectral resolution ($\Gamma\rightarrow 0$) of the Wiener-Khintchine power spectrum, the finite values of  $\Gamma$ have significant physical implications. For example, the causality of Eq.~(\ref{EWspec}) prevents the observation of the Rabi doublet before a Rabi oscillation is completed~\cite{PPR_Yamaguchi_24}. 
}

\subsubsection{The transmon artificial atom as a deformed oscillator}\label{circuit_implementation}

By choosing the function $f^2(\hat n) = 1+\alpha(\hat n -1)/2$,
the following Hamiltonian is obtained $\hat H_q = \hbar\omega_q (\hat n +1/2+\alpha\hat n^2/2)$, which we recognize as an effective Hamiltonian describing a {\it transmon} in superconducting quantum circuits \cite{BlaisRMP2021}. Here, $\hbar\omega_q = \sqrt{8E_C E_J}-E_C$, being $E_C$ and $E_J$ the charge and Josephson energies, respectively. In the transmon regime, i.e., $E_J/E_C\gg 1$, the anharmonicity factor is given by $\alpha = -E_C/\hbar\omega_q\approx -(8 E_J/E_C)^{-1/2}$. Typical experiments in circuit-QED reach values of $E_J/E_C\sim 20-50$, so one can expect anharmonicities in the range $\sim [-0.08, -0.05]$. The harmonic limit is recovered when $E_J/E_C$ takes much larger values, for which the system becomes a harmonic oscillator with frequency $\omega_q\approx \sqrt{8 E_J E_C}/\hbar$. It is well known that the weak anharmonicity of the transmon spectrum can be exploited to engineer artificial qubits, allowing only transitions between the ground and the first excited states. We can then restrict the energy levels of the transmon  to the first two levels, with energies $E_0$ and $E_1$. The resulting effective {\it transmon qubit} Hamiltonian can be written as $H_q = \hbar\tilde{\omega}_q\hat\sigma_z/2$, where $\tilde{\omega}_q=\omega_q-E_C/2\hbar$.
Therefore, the architecture of circuit-QED could allow the realization of the deformed JC model by considering two interacting anharmonic oscillators. One of these circuits will represent a Kerr oscillator, and the other one the two-level system. The nonlinear coupling might also be engineered using inductive or capacitive coupling. However, a microscopic derivation of such a circuit Hamiltonian is not straightforward. 
\subsection{Spectral response with a nonlinear coupling}\label{Section_Nonlinear_Coupling}

As in the standard Jaynes-Cummings model of Sec.~\ref{Section_2}, the total  number of excitations in the deformed Jaynes-Cummings model is constant, and we can restrict our analysis to the $n-$th doublet $\left\{\ket{e, n}, \ket{g, n + 1}\right\}$. The matrix representation of the Hamiltonian (\ref{deform_JCH}) is $H^{D}_{11}=\langle e,n|\hat H^{D}_{\texttt{JC}}|e,n\rangle$, $H^{D}_{22}=\langle g,n+1|\hat H^{D}_{\texttt{JC}}|g,n+1\rangle$, $H^{D}_{12}=\langle e,n|\hat H^{D}_{\texttt{JC}}|g,n+1\rangle=\mathrm{i}{\hbar\Omega_{n}}f(n + 1)/2$, $H^{D}_{21}=(H^{D}_{12})^*$, where
\begin{equation}
\begin{aligned}
    H^{D}_{11} &= \frac{\hbar\omega_{c}}{2}\left[n f^{2}(n) + (n + 1)f^{2}(n + 1) + 1)\right] + \frac{\hbar\Delta}{2},\\ 
    H^{D}_{22}\!&=\!\frac{\hbar\omega_{c}}{2}\!\left[(n\!+\!1)f^{2}(n\!+\!1)\!+\!(n\!+\!2)f^{2}(n\!+\!2)\!-\!1\right]\!-\!\frac{\hbar\Delta}{2}.
\end{aligned}
\end{equation}
We can rewrite the $n-$th doublet as 
\begin{equation}\label{Matrix_Elements_HJCD}
H_n^D\!=\!\frac{\hbar\omega_c}{4}\!\left(h_{11}\!+\!h_{22} \right){\mathbb{I}}_{2\times 2}^{}
\!+\!V_n,\, 
V_{n}\!=\!\frac{\hbar}{2}\left(
    \begin{array}{cc}
    \Delta_f & -\mathrm{i}\tilde \Omega_{n}\\
    \mathrm{i}\tilde \Omega_{n} & -\Delta_f
    \end{array}\right),
\end{equation}
where 
\begin{equation}
\begin{aligned}
    h_{11} &= n f^2(n) + (n + 1)f^2(n + 1) + 1,\\
    h_{22} &= (n + 1)f^2(n + 1) + (n + 2)f^2(n + 2) - 1,
\end{aligned}
\end{equation}
 $\Delta_{f,n}= \Delta +{\omega_{c}}(h_{11} - h_{22})/2$, and $\tilde{\Omega}_{n}\!=\!\Omega_{0} \sqrt{n + 1} f(n + 1)$.   

The detuning ($\Delta_{f,n}$) and the Rabi frequency ($\tilde\Omega_n$) are now functions of the deformation function; when $f(n)=1$ (no deformation), $\Delta_{f,n}\rightarrow \Delta$ and $\tilde\Omega_n\rightarrow \Omega_n$.

The energy eigenvalues of (\ref{Matrix_Elements_HJCD}) are~\cite{Cordero}
\begin{equation}\label{def_eigen}
	\mathcal{E}_{n}^{\left(\pm\right)}= \frac{\hbar\omega_{c}}{4}\left(h_{11} + h_{22}\right) \pm \frac{\hbar}{2}\sqrt{\Delta_f^2 + \tilde \Omega_{n}^{2}},
\end{equation}
with the corresponding eigenvectors
\begin{equation}\label{f_polaritons}
\begin{aligned}
    \ket{+, n} &= \cos(\theta_{n}^{D}/2)\ket{e, n} + \mathrm{i}\sin(\theta_{n}^{D}/2)|{g, n + 1}\rangle, \\
    \ket{-, n} &= \sin(\theta_{n}^{D}/2)\ket{e, n} - \mathrm{i}\cos(\theta_{n}^{D}/2)|{g, n + 1}\rangle,
\end{aligned}
\end{equation}
and with the mixing angle $\tan\theta_{n}^{D} = {\Omega}_{n}f(n+1)/\Delta_{f,n}$. We want to emphasize that, until this point, we have not given an explicit form to the deformation function and the diagonalization procedure is valid for any arbitrary choice of $f(n)$.  

\begin{figure}
    \centering
    \includegraphics[width= 0.95\linewidth, keepaspectratio]{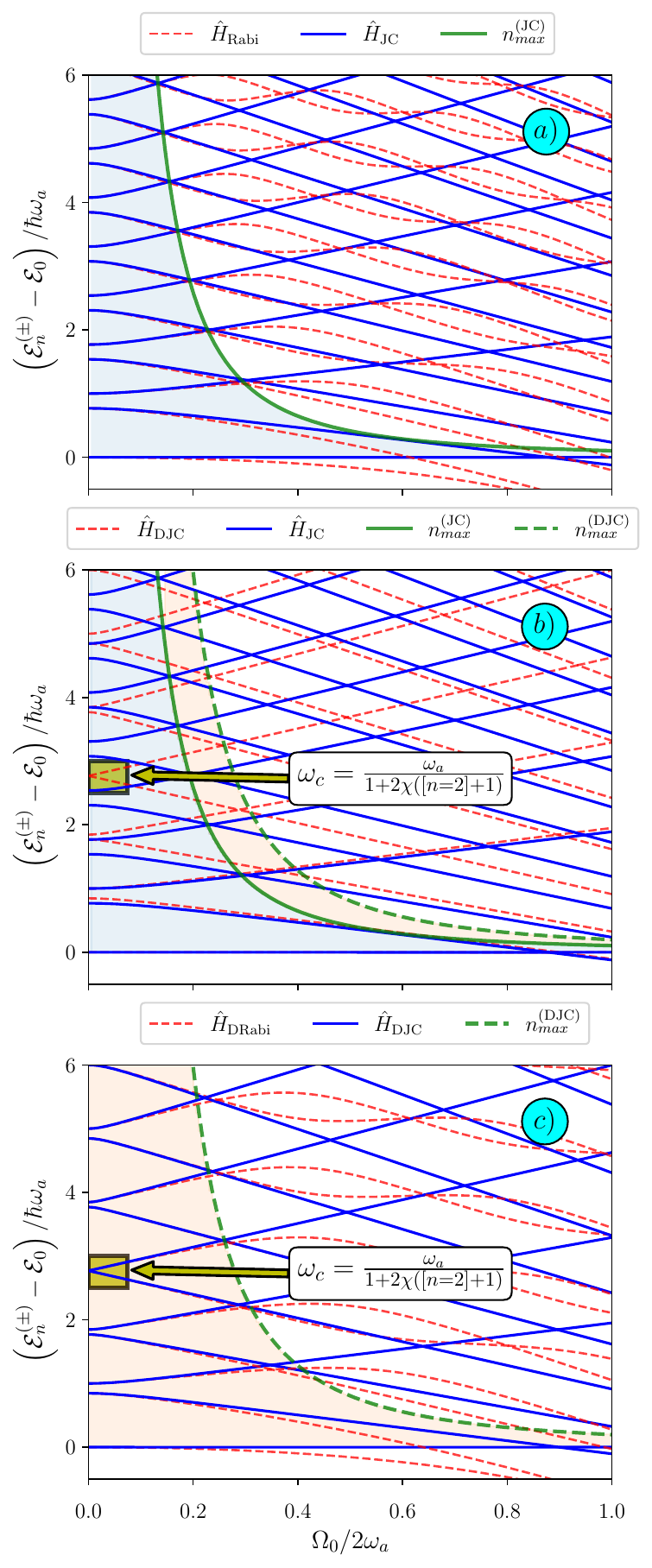}
    \caption{
    Energy levels of the $a)$ quantum Rabi Hamiltonian ($\hat{H}_{\texttt{Rabi}}$) vs. the standard JC Hamiltonian ($\hat{H}_{\texttt{JC}}$), $b)$ The deformed JC Hamiltonian, see $\hat H^{D}_{\texttt{JC}}$  in Eq.~(\ref{Deform_JC_Hamil}), vs. $\hat{H}_{\texttt{JC}}$, and $c)$ The deformed quantum Rabi Hamiltonian ($\hat{H}_{\texttt{Rabi}}^D$) vs. $\hat H^{D}_{\texttt{JC}}$.  Parameters $m=2$, $\chi = 0.05\omega_a$, and $\omega_{c}/\omega_{a} =(1 + 2\chi (m + 1))^{-1} = 0.7692$ ensure a selective transition depicted by the small rectangle in $b)$ and $c)$. The light blue and red areas indicate the RWA regime of validity, while the dashed and solid green lines are their upper bounds; see Eq.~(\ref{Limit_nmax}).} 
    \label{fig:eigenvals}
\end{figure}

Using the explicit deformation function given in Sec.~\ref{Section_Nonlinear_Oscillator} the deformed Jaynes-Cummings Hamiltonian takes the form
\begin{equation}\label{Deform_JC_Hamil}
\begin{aligned} 
     &\hat H^{D}_{\texttt{JC}} = {\hbar\omega_{c}}\Big(\hat n+\frac{1}{2}\Big)+ \hbar\omega_c\chi\Big(\hat n^2+\hat n+\frac{1}{2}\Big)+ \frac{\hbar\omega_a}{2}\hat{\sigma}_{z}\\
    & -{\rm i}\frac{\hbar}{2}\Big({\Omega_0}\sqrt{1\!+\!\chi\!+\!\chi\hat n}\Big)\hat{a}\hat{\sigma}_{+}^{} +{\rm i}\frac{\hbar}{2}\hat{a}^{\dagger}\hat{\sigma}_{-}^{}\Big({\Omega_0\sqrt{1\!+\!\chi\!+\!\chi\hat n}}\Big).
\end{aligned}
\end{equation}
Figure~\ref{fig:eigenvals}~$b)$ shows (red dashed lines), as a function of the normalized coupling constant $\Omega_0/2\omega_a$, the influence of the deformation parameter $\chi$ upon the energy eigenvalues of (\ref{Deform_JC_Hamil}) given in \eqref{def_eigen} using $f^2(n)=1+\chi n$. We see that the separation between the eigenvalues is proportional to the deformation parameter.
When no excitation is present in the field ($n=0$), the vacuum Rabi frequency still changes due to the nonlinear coupling to $\Omega_0\sqrt{1+\chi}$. As we show below, this change also appears in the two-level system's spectral response; see Eq.~(\ref{Slimit_def}). The blue solid lines correspond to the standard JC eigenvalues $E_n^{(\pm)}$ of Sec.~\ref{Section_2}. Note we choose a special condition for the ratio $\omega_c/\omega_a$ in which all bare states, states where $\Omega_0=0$, are non-degenerate except the ones having $3=2+1$ excitations, i.e., $n=2$ excitations from the cavity field and one from atom in its excited state $|e\rangle$. This situation is known as a selective transition~\cite{Cordero} and it is indicated by the small rectangle in Fig.~\ref{fig:eigenvals} $b)$ and $c)$. 

Figure~\ref{fig:eigenvals}~$c)$ shows the coupling range in which the RWA in the deformed JC Hamiltonian out of resonance ($\omega_a\neq\omega_c$), but with a selective transition ($\Delta_{f,2}=0$), can be used. We performed a numerical diagonalization, using the Python numerical suite QuTIP (Quantum Toolbox in Python) \cite{Johansson2012, Johansson2013}, of the deformed quantum Rabi Hamiltonian 
$\hat{H}_{\texttt{Rabi}}^D={\hbar}\omega_c(\hat{A}^{\dagger}\hat{A}+\hat{A}\hat{A}^{\dagger})/2+{\hbar}\omega_a\hat{\sigma}_{z}/2-\mathrm{i}\hbar{\Omega_{0}}(\hat{A}-\hat{A}^{\dagger})\hat{\sigma}_{x}/2$, and compare the corresponding eigenvalues (red dashed lines) with those of $\hat H_{\texttt{JC}}^D$ (blue solid lines). In addition, by computing the matrix element $\Omega_0^2|\langle g,n+1|\hat A \hat \sigma_{-}|e,n+2\rangle|^2/4(\omega_c+\omega_a+2\omega_c\chi(n+2))^2 \ll 1$, we obtain the upper bound for the RWA regime of validity given by
    \begin{equation}\label{Limit_nmax}
    \begin{aligned}
    n_{\rm max}^{\rm (DJC)}=&\frac{(4\chi+1)\Omega_0^2-16\chi\omega_c(\omega_a+\omega_c+4\chi\omega_c)}{2\chi(16\chi\omega_c^2-\Omega_0^2)}\\
       &\qquad\qquad\quad-\frac{\Omega_0\sqrt{\Omega_0^2+16\chi(\omega_a^2-\omega_c^2)}}{2\chi(16\chi\omega_c^2-\Omega_0^2)}.
          \end{aligned}
    \end{equation}
If $\chi\rightarrow 0$ the above equation reduced to the expression $n_{\rm max}^{\rm (JC)}=4(\omega_c+\omega_a)^2/\Omega_0^2$~\cite{Liberato2024, Solano_PRA_2017}. Figure~\ref{fig:eigenvals} displays $n_{\rm max}^{\rm (DJC)}$ ($n_{\rm max}^{\rm (JC)}$) with the dashed (solid) green line.

\textcolor{black}{While Fig.~\ref{fig:eigenvals}~ $a)$ and $c)$ compare the standard and deformed JC eigenvalues with their quantum Rabi counterparts, Fig.~\ref{fig:eigenvals}~$b)$ contrasts two Hamiltonians where the RWA has been applied. In particular, at $\Omega_0=0$, the parameter $\chi$ only modifies the bare states. Figure~\ref{fig:eigenvals}~$b)$ shows how, by increasing the energy, each doublet in the deformed JC model (red dashed lines) increases its deviation from the standard JC model (solid blue lines) until reaching a selective transition even that $\omega_c\neq\omega_a$. This transition results from the actual dependence of the nonlinear field frequency with the number of excitations, see Eq.~(\ref{EW_nonlinear_oscillator}).}

\textcolor{black}{In Appendix~\ref{app:parity}, we demonstrate that the four Hamiltonians studied in Fig.~\ref{fig:eigenvals} are invariant under the action of the parity operator $\hat{\Pi}$~\cite{Birrittella2015}. Parity preservation implies that the corresponding energy levels from dissimilar symmetry sectors can present spectral crossings \cite{Li2021}. However, when the RWA is not applied, careful inspection of Fig.~\ref{fig:eigenvals} shows that energy levels with the same parity subspace display avoided crossings; see Ref.~\cite{Braak2019} for an extensive discussion of symmetries in the quantum Rabi model and the JC model.}

\subsubsection*{EW spectrum for the atom in the deformed JC model}

To obtain the two-time autocorrelation function for the deformed JC model, we proceed in a similar form as that used for the standard JC model in Sec.~\ref{Section_2}. Setting $|\Psi(0)\rangle=\ket{e,n}$ as the initial state, we get
\begin{equation}\label{Ga}
\begin{aligned}
    &G_a(t_1,t_2) = \frac{1}{8}e^{\frac{\mathrm{i}}{\hbar}\left((E_{n}^{(0)} - E_{n-1}^{(0)})(t_{1} - t_{2})\right)}e^{-\frac{\mathrm{i}}{2}\left(\phi_{-}^{D} t_{2} + \phi_{+}^{D}t_{1}\right)}\times \\
    &\left[e^{\mathrm{i}\phi_{n}^{D} t_{1}}\left(1 + \frac{\Delta_{f, n}}{\phi_{n}^{D}}\right) + \left(1 - \frac{\Delta_{f, n}}{\phi_{n}^{D}}\right)\right] \\
    &\times\left[\left(1 + \frac{\Delta_{f, n}}{\phi_{n}^{D}}\right) + e^{\mathrm{i}\phi_{n}^{D} t_{2}} \left(1 - \frac{\Delta_{f, n}}{\phi_{n}^{D}}\right)\right]\\
    &\times\left[e^{\mathrm{i}\phi_{n - 1}^{D} (t_{1} - t_{2})}\left(1 +\frac{\Delta_{f, n - 1}}{\phi^{D}_{n - 1}}\right) + \left(1 - \frac{\Delta_{f, n - 1}}{\phi^{D}_{n-1}}\right)\right],
\end{aligned}
\end{equation}
where we have defined \textcolor{black}{$\phi_\pm^D=\phi_n^D\pm\phi_{n-1}^D$,}
\begin{equation}
\phi_n^D=\sqrt{\Delta_{f,n}^2+\tilde \Omega_n^2}
\end{equation}
 and $E_n^{(0)} = \hbar\omega_c(h_{11}+h_{22})/4$. \textcolor{black}{Furthermore, we have used the following trigonometric identities: 
$2\sin^2\left[{\arctan(x)}/{2}\right]=1-(1+x^2)^{-1/2}$ and $2\cos^2\left[{\arctan(x)}/{2}\right]=1+(1+x^2)^{-1/2}$}.

Substituting the explicit form for the deformation function (see Sec.~\ref{Section_Nonlinear_Oscillator}) in the last expression, $E_{n}^{(0)}-E_{n-1}^{(0)} = \hbar\omega_c(1+\chi(2n+1))$ and
\begin{equation}\label{Detuning_f}
\Delta_{f,n}=\omega_a\big(1-2\omega_c\chi/{\omega_a}\big)-\omega_c\big(1+2\chi n\big).
\end{equation}
Note that the resonance condition $\Delta_{f,n}=0$, known as a selective transition~\cite{Cordero}, is a function of the differences in the bare atomic and cavity frequencies, the doublet under consideration, and the deformation parameter. Note that even if $\Delta_{f,n}=0$ is satisfied in the correlation function (\ref{Ga}), $\Delta_{f,n-1}$ is not zero for $n\neq 0$, and as a consequence, the corresponding EW will be asymmetric.  

From equation~(\ref{EW_nonlinear_oscillator}) it is clear that when the field has $n$-excitations its frequency is $\omega_c(\chi,n)\equiv\omega_c(1+2\chi n)$ while the atomic one is $\omega_a(\chi)\equiv\omega_a(1-2{\omega_c}\chi/\omega_a)$ due to the nonlinear coupling. When the initial state of the system is $|e,0\rangle$, and considering the resonance condition, $\omega_a(\chi)=\omega_c(\chi,0)$ implying $\omega_c=\omega_a/(1+2\chi)$, the EW physical spectrum in the long-time limit ($\Gamma t\gg 1$) is 
\begin{equation}\label{Slimit_def}
S(\omega,\Gamma)_{\texttt{DVRS}}^{}=S_+^{}(\omega,\Gamma)+S_-^{}(\omega,\Gamma), 
\end{equation}
where $S_\pm^{}(\omega,\Gamma)= ({\Gamma}/{2})[\Gamma^2+\big(\omega-\omega_a\pm\frac{\Omega_0}{2}\sqrt{1+\chi}\big)^2]^{-1}$.
For a detuning $\Delta_{f,0}\neq 0$, $S_\pm^{}(\omega,\Gamma)$ changes to
\begin{equation}\label{Slimit_def_2}
   S_\pm^{}(\omega,\Gamma)=\Big(1\mp\frac{\Delta_{f,0}}{\phi_0^D}\Big)^2\frac{\Gamma/2}{\Gamma^2+\big(\omega-\omega_a+\frac{\Delta_{f,0}}{2}\pm\frac{\phi_0^D}{2}\big)^2},
\end{equation}
where $\Delta_{f,0}\!=\!\omega_a-\omega_c(1\!+\!2\chi)$ and $(\phi_0^D)^2\!=\!{\Delta_{f,0}^2\!+\!\Omega_0^2(1\!+\!\chi)}$.
The two peaks in the deformed vacuum Rabi splitting $S(\omega,\Gamma)_{\texttt{DVRS}}^{}$ are a clear spectral signature of the hybridization of the atomic and field quantum states, the so-called upper and lower polaritons, see Eq.~(\ref{f_polaritons}). As we are working in the context of deformed operators, we dub them {\it f}-deformed polaritons instead of the {\it q}-deformed polaritons~\cite{Harouni_JPB_2009}. The latter, as expected, will be a particular case of the former~\cite{manko1}. If $\chi=0$ and $\omega_a=\omega_c$, then $\Delta_{f,0}=0$ and $S(\omega,\Gamma)_{\texttt{DVRS}}^{}\rightarrow S(\omega,\Gamma)_{\texttt{VRS}}^{}$. 

\begin{figure}[t]
    \centering
    \includegraphics[width =\linewidth, keepaspectratio]{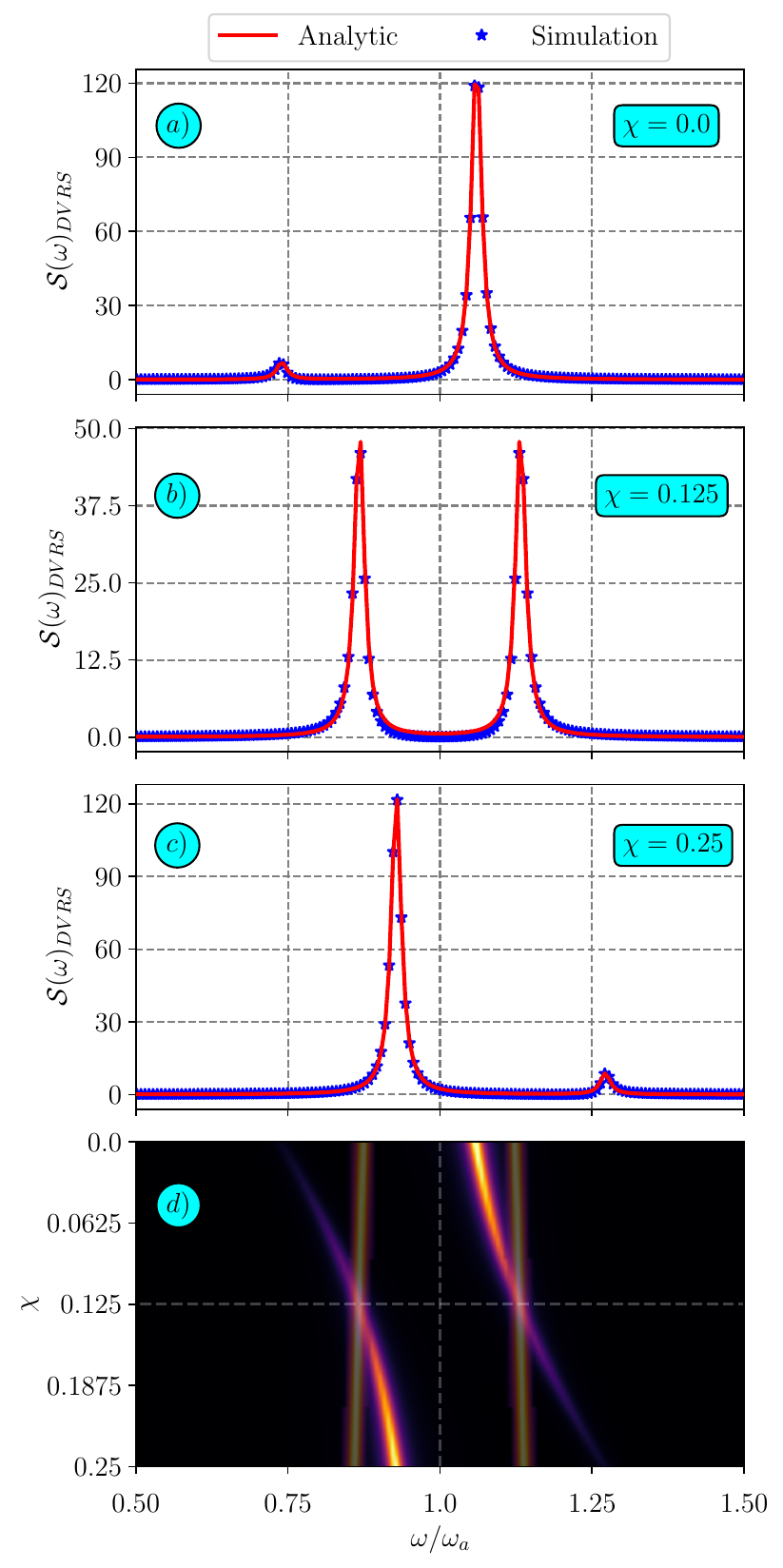}
    \caption{
    Long-time limit EW spectrum (red solid line) for the deformed vacuum Rabi splitting (DVRS), see Eq.~(\ref{Slimit_def}). Hamiltonian parameters are $\omega_{c}=0.8\omega_{a}$, $\Omega_{0} = 0.25\omega_{a}$, $\Gamma = 0.01\omega_a$, $\chi\in\left\{0, 0.125, 0.25\right\}$, see the text for details.}
    \label{fig:Long-time}
\end{figure}

We can \textcolor{black}{point out} the influence of the deformation parameter in the spectrum, looking at the long-time limit relations, where the shape of the spectrum is already defined. In Fig.~\ref{fig:Long-time} we plot (red solid line) Eq.~(\ref{Slimit_def}), for Hamiltonian parameters, $\omega_{c} =0.8\omega_{a}$, $\Gamma = 0.01\omega_a$, $\Omega_{0} = 0.25\omega_a$. 
We see a clear influence of the parameter on both, the amplitude of the spectral peaks $S(\omega,\Gamma)_{\texttt{DVRS}}$, and their position in the frequency $\omega$. a) For $\chi = 0$ the system is non-resonant, $\Delta_{f, 0} > 0$, and the spectral peaks shift to the left, indicating an asymmetry. b) For $\chi = 0.125\omega_a$, the system is resonant, $\Delta_{f, 0} = 0$, and the spectral peaks are centered around $\omega = \omega_{a}$, denoting equal amplitude. c) For $\chi = 0.25\omega_a$, the system is again non-resonant, $\Delta_{f, 0} < 0$, and the spectral peaks shift to the right, denoting once more an asymmetry. d) Density plot of the plane $\omega$~versus~$\chi$, where the constant intensity parallel lines are the $\chi$-evolution of a resonant system, $\Delta_{f, 0} = 0$ for all deformation values, note the symmetric outer drive of the lines; on the other side, the variable intensity curved lines are the $\chi$-evolution of a system moving along the deformation parameter, $\Delta_{f, 0}(\chi)$, where the crossing horizontal dashed line denote the resonant points, here the lines drive from left-to-right. Note the intensity shift between the extrema upper and lower $\chi$.
It is clear that the deformation induces a shift in the position of the peaks. 
For the case of optical cavities with a Rydberg atom in the cavity, the vacuum Rabi frequency is much smaller than the value we have used here~\cite{Haroche2006} since for other architectures yielding the same kind of Hamiltonian, for instance using superconducting elements, the Rabi and the nonlinear constant can be larger. 

\textcolor{black}{Blue stars in panels a)-c) of Fig.~\ref{fig:Long-time} denote simulation running that were obtained numerically solving the total Hamiltonian \eqref{Deform_JC_Hamil} using Python QuTip \cite{Johansson2012, Johansson2013}. The problem is set as a $N\otimes s$ size system, where $N$ and $s$ are the dimensionalities of the Hilbert space for the harmonic oscillator and the atom, respectively. In this case, $s = 2$ for a two-level system (spin-$1/2$). For cases when we have Fock number states $\ket{n}$, $N$ can be close to the value of the $n$, but for coherent states $\ket{\alpha}$, $N$ must be escalated according to the spread in the photon number distribution of $\alpha$. A propagator $U(t)$ is obtained using Adams multistep ODE integration \cite{Hairer2009-vt} of polynomial order $12$ and requiring $10^{4}$ calls between steps (iterations), returning a set of matrices evaluated in the sampled time space $[0, t_{f}]$, whose size $d$ is of order $\approx 4(2\pi/\Omega_{n})$ to avoid numerical errors. We devised an ad-hoc numerical integration for the physical spectrum in Eq.~\eqref{EWspec}, obtaining the correlation functions as matrices evaluated at $t_{i}$ in the trace representation, and a trapezoidal integration of the integrals gives the final form of the spectrum. We present a reproducible repository with the numerical simulation implementation in~\cite{EW_JC_repo}.} 
It is worth mentioning that the deformed VRS (\ref{Slimit_def}) captures the energy transitions from the eigenstates $|\pm,0\rangle$ to the ground state of the system, and since the ground state is decoupled due to the RWA, the resulting EW spectrum is the only symmetric spectral response of this nonlinear JC model. As shown below, this is not the case when considering an initial excitation of the nonlinear field; we show how the EW spectrum is asymmetric even for a selective transition~\cite{Cordero}.
%
\begin{figure}[h]
    \centering
    \includegraphics[scale=0.5]{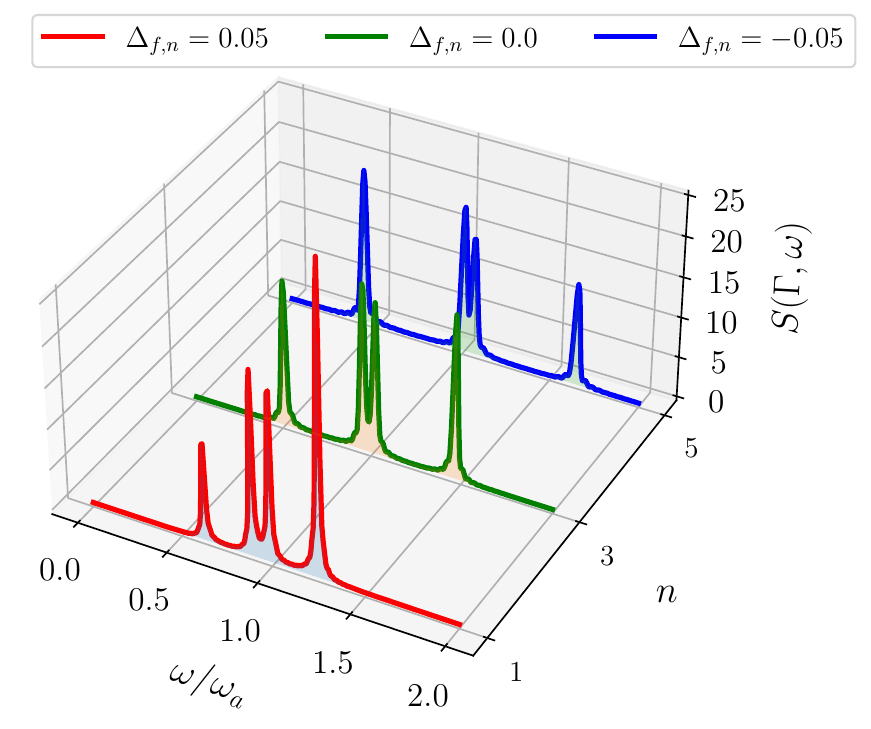}
    \caption{The EW spectrum for the two-level atom as a function of the number of photons $n$ in the nonlinear field, for fixed $\chi$. Note that $n$ leads to a sign change in $\Delta_{f,n}$, denoting a spread in the frequency $\omega$ and a transition in the asymmetry of the spectral peaks; see Eq.~(\ref{Detuning_f}). Hamiltonian parameters: $\omega_{c} = 0.9\omega_{a}$, $\Omega_{0} = 0.25\omega_{a}$, $\Gamma = 0.01\omega_a$, $t = 16\tau(4)$, $n\in\left\{1, 3, 5\right\}$}    
    \label{fig:EW_chievolution_number_1}
\end{figure}
\begin{figure}[h]
    \centering
    \includegraphics[scale=0.5]{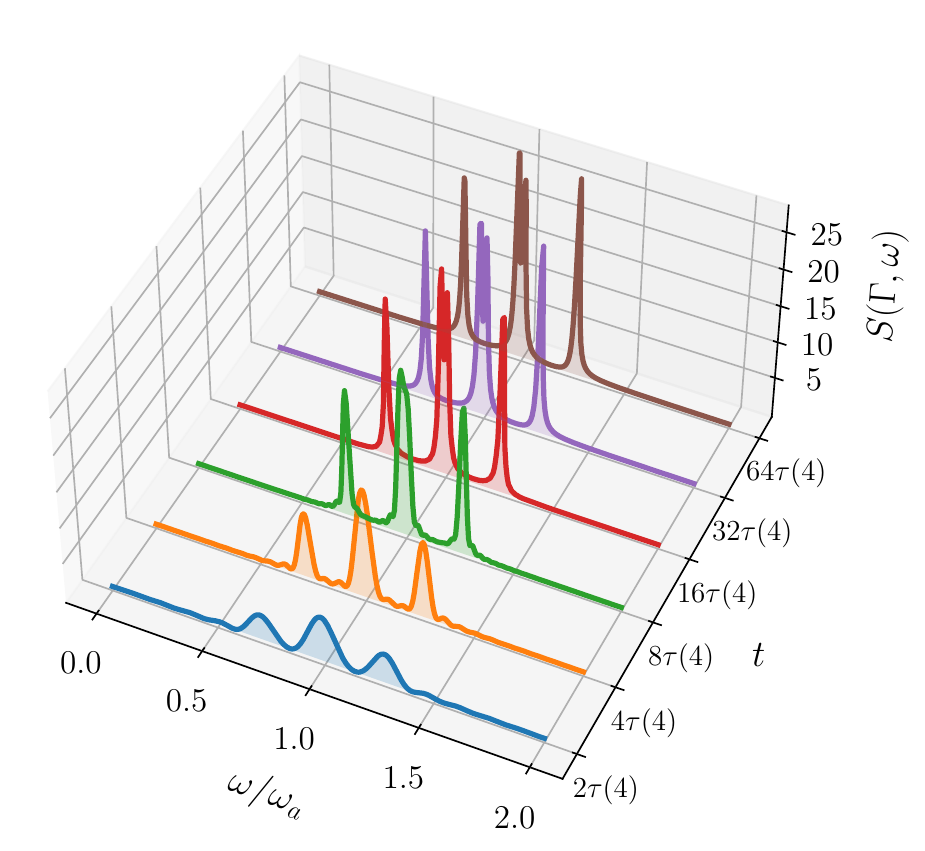}
    \caption{
    Temporal evolution of the EW spectrum for the atom with an initial state $|\Psi(0)\rangle = |e,n\rangle$. Hamiltonian parameters: $n=4$, $\chi = 0.0125\omega_a$, $\omega_{c} = \omega_{a}/(1 + 2 \chi (n + 1))$, $\Omega_{0} = 0.125 \omega_a$, $\Gamma = 0.01 \omega_a$, $t\in\left\{2.0, 4.0, 8.0, 16.0, 32.0, 64.0\right\}\tau(4)$}
    \label{fig:EW_chievolution_number_2}
\end{figure}

In figure~\ref{fig:EW_chievolution_number_1}, we show the EW spectrum for the two-level atomic system with an initial state $|\Psi(0)\rangle = |e, n\rangle$ as a function of the photon number $n$ with fixed deformation $\chi$. Notice that the detuning given by Eq.~(\ref{Detuning_f}) is a function of the photon number and the deformation parameter, then variations in the photon number are equivalent to a modification of the deformation. In the figure, we show the effect when there is a sign change in $\Delta_{f,n}$ due to the change of photon number. There is an asymmetry in the heights of the peaks; for positives $\Delta_{f,n}$, the peaks with lower frequency are smaller, and the reverse is true for negative $\Delta_{f,n}$. For this calculation, we used Eq.~(\ref{Ga}) and 16 periods for the evolution, given by the Rabi period $\tau(n) = 2\pi/\Omega_{n}$. We see a similar behavior as that noted in the DVRS shown in Fig~\ref{fig:Long-time} but with an amplitude shift of the spectral peaks, plus an evident displacement in the position of the peaks. However, in this case, even for the selective transition ($\Delta_{f,3}=0$), the corresponding spectral response is slightly asymmetric, a clear signature for the impossibility of getting resonant conditions for all bare states; see Fig.~\ref{fig:eigenvals}. 

Figure~\ref{fig:EW_chievolution_number_2} shows the temporal evolution of the EW spectrum for a fixed number state $n=4$ and deformation parameter $\chi=0.0125\omega_a$, with $\Omega_0 = 0.125\omega_a$ and $\omega_a = 1$ under a selective transition $\Delta_{f,4}=0$. We have chosen a set of specific times, which are multiples of the period $\tau(n)$. Notice how the peaks are being formed as the system is evolved to longer times, in this calculation we used the expressions we got for the EW spectrum in the most general case (too cumbersome to show here), obtained from the two-time autocorrelation function given by Eq.~(\ref{Ga}). We mention that our analytical result was contrasted with a purely numerical calculation and mutually agree, since our expression for the two-time autocorrelation function is exact. 
\begin{figure}[b]
    \centering
    \includegraphics[scale=0.5]{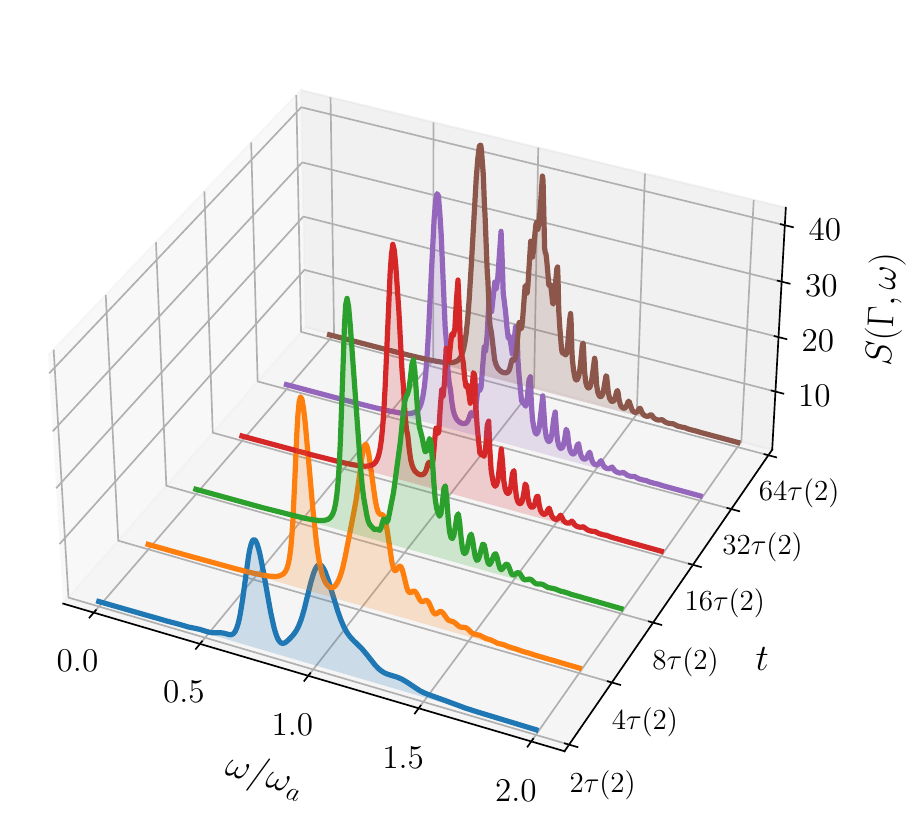}
    \caption{
    Temporal evolution of the EW spectrum for the atom with an initial state $|\Psi(0)\rangle =|e,\alpha\rangle$. Hamiltonian parameters: $\chi = 0.0125$, $\omega_{a} = 1.0$, $\omega_{c} = \omega_{a}/(1 + 2 \chi (n + 1))$, $\Omega_{0} = 0.125 \omega_a$, $\Gamma = 0.01 \omega_a$, $\alpha = 2$, $t\in\left\{2.0, 4.0, 8.0, 16.0, 32.0, 64.0\right\}\tau(2)$}
    \label{fig:EW_timeevolution_coherent}
\end{figure}

The relevance of the initial conditions for the EW spectrum is shown in Fig.~\ref{fig:EW_timeevolution_coherent} where we consider the temporal evolution of the spectrum for an initial state $|\Psi(0)\rangle =|e\rangle\otimes|\alpha\rangle$ where $|\alpha\rangle$ is a coherent state for the cavity field and the atomic system is in the excited state.  
Notice that the resonance condition depends upon the number state of the field, then only one transition is at resonance for a given value of the deformation parameter and $n$. At the initial time, the spectrum consists of two broad peaks and as the evolution is being carried out, we notice the appearance of multiple peaks due to the multiple transitions allowed. In the case shown in the figure, we used a coherent state with an average photon number of $\bar n=4$ and $\Omega_0=0.125\omega_a$, $\chi=0.0125\omega_a$. 

\begin{figure}[t]
    \centering
    \includegraphics[scale=0.5]{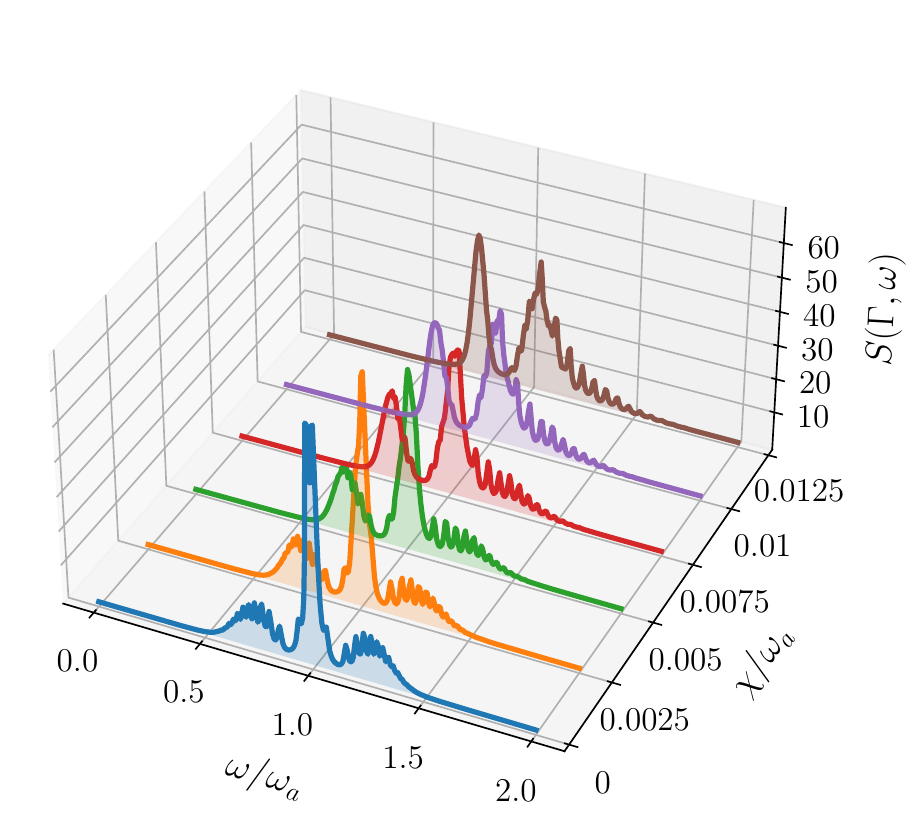}
    \caption{
    EW spectrum for the atom as a function of the deformation parameter $\chi$ with initial state $|\Psi(0)\rangle = |e,\alpha\rangle$. Hamiltonian parameters: $\omega_{a} = 1.0$, $\omega_{c} = \omega_{a}/(1 + 2 \chi (n + 1))$, $\Omega_{0} = 0.125\omega_a$, $\Gamma = 0.01 \omega_a$, $\alpha = 2$, $t = 16\tau(2)$, $\chi\in\left\{0, 0.0025, 0.005, 0.0075, 0.01, 0.0125\right\}$}
    \label{fig:EW_chievolution_coherent}
\end{figure}

Figure~\ref{fig:EW_chievolution_coherent} shows the EW spectrum for the atom as a function of the deformation parameter $\chi$ in the range $0\leq\chi/\omega_a\leq 0.0125$. We see that when there is no deformation, the spectrum is symmetric, with two peaks at the field's frequency $\omega_c/\omega_a=1$ and two broad packets with multiple transitions located at $\omega_c\pm \Delta\omega$ with $\Delta\omega\simeq 0.5$; see also~\cite{Castro_PRA_1996}. When the deformation is different from zero, we notice an asymmetry in the packets, the intensity of the central peaks decreases and for larger values of $\chi$ the spectrum consists of two broad packets with multiple transitions each. 

\textcolor{black}{
\subsubsection{Physical implementation of the nonlinear JC model}
}
\textcolor{black}{
In Sec.~\ref{circuit_implementation} we have shown how to implement the nonlinear field (nonlinear oscillator) in superconducting circuits. On the other hand, it is known that the nonlinear (intensity-dependent) coupling between an oscillator and a two-level system naturally emerges in the physics of trapped ions. For example, references~\cite{Vogel,Moya_Reports12,PRA_Solano_18} deduced the nonlinear interaction $H_{\rm nJC}={i}g[\hat\sigma_+f(\hat n)\hat a-\hat\sigma_-\hat a^\dagger f(\hat n)]$, where $f(\hat n)$ is given by Eq.~(\ref{f_ions}).
At least two methods are known for physically implementing a deformed oscillator and a deformed coupling simultaneously. One approach is in semiconductor physics, where excitons interacting with a single radiation mode of a microcavity are described by a Hamiltonian that includes Kerr-type nonlinearities and an intensity-dependent coupling with the field~\cite{Liu_PRA_2001,Harouni_JPB_2009}. This semiconductor realization occurs in quantum dots (QDs) because the corresponding exciton creation operator changes its statistic from a Bose-Einstein (for large QDs) to a Fermi-Dirac (for small QDs) due to the Pauli exclusion principle~\cite{Laussy_PRB_2006}. 
The second realization of the nonlinear JC model can be found in optics, specifically in integrated photonics, where extensive research has been conducted by simulating quantum systems using photonic lattices~\cite{PRL_Moya_11,Perez_Bloch}. In this photonic architecture, the probability amplitude evolution of a given quantum system is analogous to the spatial light propagation in arrays of high-quality coupled optical waveguides. These waveguides are inscribed in silica glass using femtosecond-laser writing technology and feature customized transverse refractive index and nonuniform coupling coefficients~\cite{Szameit_2010}. 
The photonic realization of the linear JC model was proposed in reference~\cite{OL_Longhi_11}, and shortly after, the experimental photonic realization of the quantum Rabi model was implemented~\cite{PRL_Longhi_12}. The nonlinear (deformed) JC model and the nonlinear (deformed) quantum Rabi model were proposed for implementation in photonic lattices using experimentally feasible values in~\cite{Blas_OpExp_2013}. However, the corresponding spectral responses were not explored.
}

\section{Conclusions}\label{Section_4}

Using the time-dependent physical spectrum of Eberly and W\'odkiewicz, we derived the spectroscopy of a generalized Jaynes-Cummings model, in which nonlinearities in the cavity field and the coupling were introduced via f-deformed operators. Interestingly, we showed that the nonlinear cavity field has a spectral response similar to that of a superconducting qubit operating at the strong dispersive regime of circuit-QED, demonstrating its self-interaction (Kerr) nature from a purely spectroscopic standpoint; see Fig.~\ref{EW_Nonlinear_Field_2}. Since the nonlinear cavity field frequency strongly depends on its initial state, getting the usual resonant condition with the two-level atomic system is  not possible for finite field excitations. When considering the nonlinear coupling, the atomic spectral response is always asymmetric for all used parameters except for the vacuum Rabi splitting; see Figs.—\ref{fig:Long-time} to \ref{fig:EW_chievolution_coherent}. We also calculated the upper bound for the validity of the rotating-wave approximation in the nonlinear JC model, corroborating it with a purely numerical solution of the nonlinear quantum Rabi model, see Fig.~\ref{fig:eigenvals}. By providing this detailed spectral analysis, we hope our results can be of interest to the quantum optics community working on f-deformed and q-deformed systems by inspiring the search for their spectral responses, which are generally overlooked in those approaches. 

\acknowledgments
R.R.-A. thanks DGAPA-UNAM, Mexico for support under Project No. IA104624. A.R.-U. acknowledges DGAPA-UNAM, Mexico for support under Posdoc-DGAPA 2023-2024 program, and to ICF-UNAM for the in-place support, J.R, L.M-D and D.A.-L acknowledge partial support from DGAPA-UNAM project IN109822. We acknowledge support from CONAHCYT-Mexico under grant CBF2023-2024-2888.  

\appendix

{\color{black}
\section{Autocorrelation function and the EW spectrum of the JC model}\label{app:corrfunc}
We present a detailed description of the methodology used to obtain the two-time autocorrelation function for the atom, \eqref{func:acorr}, defined as
\begin{equation}\label{eq:fcorr}
\begin{aligned}
G(t_{1},\!t_{2})_{\rm atom}^{}\!=\! 
    \braket{\Psi(0)\vert\hat{U}^{\dagger}\!(t_{1}\!)\hat{\sigma}_{+}\hat{U}(t_{1}\!) \hat{U}^{\dagger}(t_{2}\!)\hat{\sigma}_{-}\hat{U}(t_{2}\!)\vert\Psi(0)}\!, 
\end{aligned}
\end{equation}
where $\hat{U}(t_j)=\exp(-{\rm i} \hat H_{\texttt{JC}}t_j/\hbar)$ is the time-evolution operator of the JC model.
We first recall the diagonalization of the Jaynes-Cummings Hamiltonian 
\begin{equation}
    2\hat H_{\texttt{JC}} = \hbar\omega_c(\hat{a}^{\dagger}\hat{a}+\hat{a}\hat{a}^{\dagger})+\hbar{\omega_a}\hat \sigma_z - \mi\hbar{\Omega_0}(\hat a \hat{\sigma}_{+}-\hat a^{\dagger}\hat{\sigma}_{-}),
\end{equation}
without the zero-point energy $\hbar\omega_c/2$ it can be written as
\begin{equation}
    {H}_n = \hbar\omega_{c}\big(n+{1}/{2}\big){\mathbb{I}}_{2\times 2}^{}+V_{n},\,\,V_{n} = \frac{\hbar}{2}\begin{pmatrix}
		\Delta & -\mathrm{i}\Omega_{n}\\
		\mathrm{i}\Omega_{n} & -\Delta
	\end{pmatrix}\!,
\end{equation}
where ${\mathbb{I}}$ is the identity matrix, $\Delta = \omega_{a} - \omega_{c}$ is the atom-field detuning, and $\Omega_n=\Omega_0\sqrt{n+1}$ is the $n$-photon Rabi frequency.
Diagonalizing $V_n$ we obtain the eigenvectors
\begin{equation}
\begin{aligned}
    |+,n\rangle &= \cos{\left(\frac{\theta_n}{2}\right)}|e,n\rangle + \mi\sin{\left(\frac{\theta_n}{2}\right)}|g,n+1\rangle, \\
    |-,n\rangle &= \sin{\left(\frac{\theta_n}{2}\right)}|e,n\rangle-\mi\cos{\left(\frac{\theta_n}{2}\right)}|g,n+1\rangle ,
\end{aligned}
\end{equation}
and the eigenvalues $E_n^{(\pm)}=\hbar\omega_c(n+1/2)\pm \frac{\hbar}{2}\sqrt{\Delta^2+\Omega_n^2}$,
where the mixing angle is $\tan{\theta_n} = \Omega_n/\Delta$. Reducing under resonance conditions, $\Delta=0$, $\theta_n=\pi/2$ and $E_n^{(\pm)}=\hbar\omega_a(n+1/2)\pm\hbar\Omega_n/2$. Then, the dressed and the uncoupled states are related by
\begin{equation}
    |\pm, n\rangle = \left( |e,n\rangle \pm \mi |g,n+1\rangle \right)/\sqrt{2}.
\end{equation}
Assuming that the initial state of the system is
\begin{equation}
    |\Psi(0)\rangle = |e,n\rangle =\left(|+,n\rangle +|-,n\rangle \right)/\sqrt{2}
\end{equation}
we apply the first unitary transformation given in~({\ref{eq:fcorr}}) to the initial state
\begin{equation}
\begin{aligned}
    &\hat{U}(t_2)|e,n\rangle  =  \\
    &\hspace{2.em}\frac{e^{-\mi\omega_a(n+1/2)t_2}}{\sqrt{2}}\left(e^{-\mi{\Omega_n t_2}/{2}}|+,n\rangle + e^{{\rm i}{\Omega_n t_2}/{2}}|-,n\rangle \right) \\
    & = e^{-\mi\omega_a(n+1/2)t_2}\times\\
    &\hspace{2.5em}\left[\cos{\left(\frac{\Omega_n t_2}{2}\right)} |e,n\rangle -\sin{\left(\frac{\Omega_n t_2}{2}\right)}|g,n+1\rangle\right],
\end{aligned}
\end{equation}
 we now apply the operator $\hat \sigma_{-}$,
\begin{equation}
    \hat{\sigma}_{-} \hat{U}(t_2)|e,n\rangle = e^{-\mi\omega_a(n+1/2)t_2}\cos{\left(\frac{\Omega_n t_2}{2}\right)} |g,n\rangle
\end{equation}
and go to the coupled basis to apply the second transformation
\begin{equation}
\begin{aligned}
    &\hat{U}^{\dagger}(t_2)\hat{\sigma}_{-}\hat{U}(t_2)|e,n\rangle = -\mi\frac{e^{-\mi\omega_a t_2}}{\sqrt{2}}\cos{\left(\frac{\Omega_n t_2}{2}\right)}\\
    &\times\left[e^{\mi\frac{\Omega_{n-1}}{2}t_2}|+,n-1\rangle - e^{-\mi\frac{\Omega_{n-1}}{2}t_2}|-,n-1\rangle \right].
\end{aligned}
\end{equation}
Applying the operator $\hat{U}(t_1)$ to this result yields
\begin{equation}
\begin{aligned}
    &\hat{U}(t_1)\hat{U}^{\dagger}(t_2)\hat{\sigma}_{-}\hat{U}(t_2)|e,n\rangle =\\
    &-\mi\frac{e^{-i\omega_a t_2}}{\sqrt{2}} e^{-\mi\omega_a (n-1/2) t_1}\cos{\left(\frac{\Omega_n t_2}{2}\right)}\\
    &\times \left[e^{-\mi\frac{\Omega_{n-1}}{2}(t_1-t_2)}|+,n-1\rangle -e^{\mi\frac{\Omega_{n-1}}{2}(t_1-t_2)}|-,n-1\rangle  \right].
\end{aligned}
\end{equation}
Now transform to the uncoupled basis and apply $\hat \sigma_{+}$, the result is
\begin{equation}
\begin{aligned}
    \hat\sigma_{+}\hat{U}(t_1)&\hat{U}^{\dagger}(t_2)\hat{\sigma}_{-}\hat{U}(t_2)|e,n\rangle = \\
    & \cos{\left(\frac{\Omega_n t_2}{2}\right)} \cos{\left(\frac{\Omega_{n-1}}{2}(t_1-t_2)\right)}\times\\ 
    &\hspace{7.5em} e^{-\mi\omega_a t_2}e^{-\mi\omega_a(n-1/2) t_1} |e,n\rangle.
\end{aligned}
\end{equation}
Finally, apply the operator $\hat{U}^{\dagger}(t_1)$ to this result and project upon $\langle e,n|$ 
\begin{equation}
\begin{aligned}
    &\langle e,n|  \hat{U}^{\dagger}(t_1) \hat{\sigma}_{+}\hat{U}(t_1)\hat{U}^{\dagger}(t_2)\hat{\sigma}_{-}\hat{U}(t_2)|e,n\rangle = \\
    &\cos{\left(\frac{\Omega_n}{2}t_2\right)}\cos{\left(\frac{\Omega_{n-1}}{2}(t_1-t_2)\right)}\cos{\frac{\Omega_n}{2}t_1} e^{\mi\omega_a(t_1-t_2)}
\end{aligned}
\end{equation}
so that
\begin{equation}
\begin{aligned}
    G(t_1,t_2)_{\texttt{atom}} & =  e^{\mi\omega_a(t_1-t_2)} \cos{\left(\frac{\Omega_n t_2}{2}\right)} \\ &\times\cos{\left(\frac{\Omega_{n-1}}{2}(t_1-t_2)\right)}\cos{\left(\frac{\Omega_n t_1}{2}\right)}, 
\end{aligned}
\end{equation}
which is Eq.~\eqref{func:atom_corr} in the main text.
When $n=0$ the correlation function is
\begin{equation}
G(t_1,t_2)_{\texttt{atom}} = e^{\mi\omega_a(t_1-t_2)}\cos{\left(\frac{\Omega_0 t_1}{2}\right)} \cos{\left(\frac{\Omega_0 t_2}{2}\right)}
\end{equation}
and then, the EW spectrum (\ref{EWspec}) has the form
\begin{equation}
\begin{aligned}
    S(\omega,\Gamma,t) &= 2\Gamma e^{-2\Gamma t} \int_0^t dt_1\, e^{(\Gamma-\mi(\omega-\omega_a))t_1}\cos{\left(\frac{\Omega_0 t_1}{2}\right)}\times \\ 
    &\phantom{= 2\Gamma e^{-2\Gamma t}}\int_0^{t} dt_2\, e^{(\Gamma+\mi(\omega-\omega_a))t_2}\cos{\left(\frac{\Omega_0 t_2}{2}\right)}\\
    & = 2\Gamma e^{-2\Gamma t} \Bigg\vert\int_0^t d\tau\, e^{(\Gamma-\mi(\omega-\omega_a))\tau}\cos{\left(\frac{\Omega_0\tau}{2}\right)}\Bigg\vert^{2},
\end{aligned}
\end{equation}
where each one of the integrals can be done analytically, and as a result, we obtain the explicit expression for the spectrum
\begin{widetext}
\begin{equation}
    \begin{aligned}
         S(\omega,\Gamma,t) &= 
         \frac{\Gamma}{2}\left[\frac{1-2e^{-\Gamma t}\cos(\omega-\omega_a+\frac{\Omega_0}{2})+e^{-2\Gamma t}}{\Gamma^2+(\omega-\omega_a+\frac{\Omega_0}{2})^2}\right.
         + \frac{1-2e^{-\Gamma t}\cos(\omega-\omega_a-\frac{\Omega_0}{2})+e^{-2\Gamma t}}{\Gamma^2+(\omega-\omega_a-\frac{\Omega_0}{2})^2}
        \\ & \hspace{4.9cm}+\left. 2 \mathrm{Re}\left\lbrace \frac{e^{i \Omega_0 t}-e^{- \Gamma t}\left( e^{\mi(\omega-\omega_a+\frac{\Omega_0}{2})t}+e^{-\mi(\omega-\omega_a-\frac{\Omega_0}{2})t}\right)+e^{-2\Gamma t}}{(\Gamma+\mi(\omega-\omega_a+\frac{\Omega_0}{2}))(\Gamma-\mi(\omega-\omega_a-\frac{\Omega_0}{2}))}\right\rbrace\right].
    \end{aligned}
\end{equation}
\end{widetext}
Taking the long-time limit $\Gamma t\gg 1$, keeping only the leading terms we get~\cite{Sanchez_PRL_1983, H_Paul_1986}
\begin{equation}
S(\omega,\Gamma) = \frac{\frac{\Gamma}{2}}{\Gamma^2+(\omega-\omega_a+\frac{\Omega_0}{2})^2} + \frac{\frac{\Gamma}{2}}{\Gamma^2 +(\omega-\omega_a-\frac{\Omega_0}{2})^2}
\end{equation}
that is Eq.~(\ref{Slimit_nondef}) in the main text.

\section{Parity preservation of the nonlinear JC model}\label{app:parity}
	The parity operator in $\mathbb{Z}_{2}$ is $\hat\Pi=\exp({\rm i}\pi\hat N_{\rm exc})$~\cite{Birrittella2015, Braak2019}, where $\hat N_{\rm exc}=\hat a^\dagger\hat a+\hat\sigma_+\hat\sigma_-$ is the excitation number operator and it can be written as
	\begin{equation}
		\hat{\Pi} = \exp\lbrace \mi\pi[\hat{n} +{\left(\hat{\sigma}_{z} + 1\right)}/{2}]\rbrace=-\hat{\sigma}_{z}\exp\left(\mi\pi\hat{n}\right).
	\end{equation}
	The condition of parity preservation $\hat{\Pi}^{\dagger}\, \hat{H}_{j}\, \hat{\Pi} = \hat{H}_{j}$,
	for $j = \texttt{JC},\,\, \texttt{DJC},\,\, \texttt{Rabi},\,\, \textrm{or}\,\, \texttt{DRabi}$ must be fulfilled. In each of the four Hamiltonians, the free evolution terms are invariant, since they appear as functions of $\{ \hat n, f(\hat n), \hat{\sigma}_{z}\}$. The action of the parity operator on the non-deformed bosonic field operators yields
	\begin{equation}
		\hat{\Pi}^{\dagger}\, \hat{a}\, \hat{\Pi} = \hat{a}e^{-\mi\pi},\quad \hat{\Pi}^{\dagger}\, \hat{a}^{\dagger}\, \hat{\Pi} = \hat{a}^\dagger e^{\mi\pi},.
	\end{equation}
	For the deformed bosonic field operators we have 
	\begin{equation}
		\hat{\Pi}^{\dagger}\, \hat{A}\, \hat{\Pi} = \hat{A}e^{-\mi\pi},\quad \hat{\Pi}^{\dagger}\, \hat{A}^{\dagger}\, \hat{\Pi} = \hat{A}^\dagger e^{\mi\pi},
	\end{equation}
where we used the commutation relations $[\hat{n}, \hat{A}^{\dagger}] = \hat{A}^{\dagger}$, and $[\hat{n}, \hat{A}] = -\hat{A}$. For the atomic operators we get
\begin{equation}
    \hat{\Pi}^{\dagger}\, \hat{\sigma}_\pm \hat{\Pi} = \hat{\sigma}_{\pm}e^{\pm\mi\pi}, \quad
    \hat{\Pi}^{\dagger}\, \hat{\sigma}_{x}\, \hat{\Pi} = -\hat{\sigma}_{x}.
\end{equation}
	Thus, we can see the invariance in each Hamiltonian. For the interaction term of the JC model and the deformed JC model we have
	\begin{equation}
	\begin{aligned}
			\hat{\Pi}^{\dagger}\, \left(\hat{a}\hat\sigma_{+} - \hat{a}^{\dagger}\hat{\sigma}_{-}\right)\, \hat{\Pi}
			&= \hat{a}\hat\sigma_{+} - \hat{a}^{\dagger}\hat{\sigma}_{-},\\
			\hat{\Pi}^{\dagger}\, (\hat{A}\hat\sigma_{+} - \hat{A}^{\dagger}\hat{\sigma}_{-})\, \hat{\Pi}
			&= \hat{A}\hat\sigma_{+} - \hat{A}^{\dagger}\hat{\sigma}_{-}.
	\end{aligned}
	\end{equation}
	For the  quantum Rabi model and the deformed quantum Rabi model we obtain
	\begin{equation}
	\begin{aligned}
		\hat{\Pi}^{\dagger}\, [\hat{\sigma}_{x}(\hat{a} - \hat{a}^{\dagger})]\, \hat{\Pi}
		&= -\hat\sigma_{x}(\hat{a}e^{-\mi\pi} - \hat{a}^{\dagger}e^{\mi\pi})\\
		&= -\hat\sigma_{x}(-\hat{a} + \hat{a}^{\dagger})\\
		&= \hat{\sigma}_{x}(\hat{a} - \hat{a}^{\dagger}),\\
		\hat{\Pi}^{\dagger}\, [\hat{\sigma}_{x}(\hat{A} - \hat{A}^{\dagger})]\, \hat{\Pi}
		&= -\hat\sigma_{x}(\hat{A}e^{-\mi\pi} - \hat{A}^{\dagger}e^{\mi\pi})\\
		&= -\hat\sigma_{x}(-\hat{A} + \hat{A}^{\dagger})\\
		&= \hat{\sigma}_{x}(\hat{A} - \hat{A}^{\dagger}).
	\end{aligned}
	\end{equation}
	These transformations demonstrate that the four Hamiltonian are invariant under the action of the parity operator $\hat{\Pi}$. 
 }

%

\end{document}